\documentclass[draftclsnofoot,onecolumn]{IEEEtran}  
\IEEEoverridecommandlockouts
\usepackage{amsmath,amssymb,amsfonts,bm}
\usepackage{algorithm,algorithmic}
\usepackage{array}
\usepackage{setspace}
\usepackage{subfig}
\usepackage{textcomp}
\usepackage{stfloats}
\usepackage{url}
\usepackage{xcolor}
\usepackage{verbatim}
\usepackage{graphicx}
\usepackage{cite}
\usepackage{makecell}
\usepackage{comment}
\usepackage{epstopdf}
\usepackage{booktabs}
\usepackage{mathrsfs}
\usepackage{bbding}
\usepackage{pifont}
\usepackage{threeparttable}
\usepackage{float}
\usepackage{caption}
\usepackage{multirow}

\def\BibTeX{{\rm B\kern-.05em{\sc i\kern-.025em b}\kern-.08em
    T\kern-.1667em\lower.7ex\hbox{E}\kern-.125emX}}

\begin{document}
\title{Reliable Clutter Suppression for Slow-Moving Weak Target Radar Detection}

\author{Ruihang Zhang, Jiayin Xue, Tingting Zhang,~\IEEEmembership{Member,~IEEE}}

\markboth{Journal of \LaTeX\ Class Files,~Vol.~14, No.~8, August~2021}%
{Shell \MakeLowercase{\textit{et al.}}: A Sample Article Using IEEEtran.cls for IEEE Journals}

\IEEEpubid{0000--0000/00\$00.00~\copyright~2021 IEEE}

\maketitle

\doublespacing      
\begin{abstract}
Reliable slow-moving weak target detection in complicated environments is challenging due to the masking effects from the surrounding strong reflectors. The traditional Moving Target Indication (MTI) may suppress the echoes from not only the static interference objects (IOs), but also the desired slow-moving weak target. According to the low-rank and sparse properties of the range-velocity maps across different radar scans, a novel clutter suppression scheme based on the Go decomposition (Godec) framework is proposed in this paper. The simulation results show that with the existence of masking effects, the target detection scheme based on Godec clutter suppression can reliably detect the slow-moving weak target, compared to the traditional MTI-based scheme. Besides, the time consumption comparison is conducted, demonstrating that the proposed solution is one that sacrifices time complexity in exchange for enhanced reliability. Additionally, the tradeoffs among the number of false alarm cells, the detection probability and the iteration times for convergence have been revealed, guiding parameter settings of the proposed solution in practical applications. Experiment validation is also conducted to verify the proposed solution, providing further insight into the scenarios where the solution is most applicable.
\end{abstract}


\section{Introduction}
The rapid development of the low-altitude economy has promoted the prosperity of the unmanned aerial vehicles (UAVs) in the daily life, which also brings many troubles at the same time, such as unauthorized surveying and mapping, interference in civil aviation, aerial photography leaks, etc. \cite{Wang2021uavrisk}. The application of integrated sensing and communication enables the base stations to position the surrounding objects while communicating with the mobile users, providing the prerequisite on the infrastructure for detection of targets in civil areas \cite{wangisacpre2024}. \par
Slow-moving weak targets, like the UAVs mentioned, are the objects that have small radar cross sections (RCS) with low relative velocities \cite{Ziwen2022kt}. Reliable detection of such targets is attractive in many civil and military applications. One challenge is that in complicated environments, the echoes of the surrounding strong reflectors usually lead to miss detection of those targets of interest (TOIs) \cite{Su2021rpca}. Various researches have been carried out recently to improve the detection of the slow-moving weak targets. Most of them concentrate on the methods of long-time coherent accumulation of the received signals, e.g., the Keystone transform, the Radon fourier transform, the fractional fourier transform, etc. \cite{Jiahao2022kt,Niu2019kt,Zhenyuan2022rft,Xiaolong2023frft,Shi2020stfrft}. Besides, there are also some of them concentrating on enhancing the detection performance of the constant false alarm rate (CFAR) detectors, such as the compressed sensing CFAR detector, the ordered statistic of sub-reference cells CFAR detector, the iterative threshold segmentation CFAR detector, etc. \cite{cao2020cscfar,Jeong2022complexity,Yang2023itscfar,Xu2017improvedsocacfar}. However, there is another trend to further improve the detection performance by suppressing the strong clutter of the interference objects (IOs), which is also the main focus in this paper. \par
The traditional method of clutter suppression like Moving Target Indication (MTI) makes full use of the difference between the static and moving objects in the Doppler domain \cite{Xing2022mti}. The authors in \cite{skolnik1980introduction} and \cite{Ash2018mti} respectively introduced the application of MTI for pulsed and linear frequency-modulated continuous wave (LFMCW) radar, for moving target detection. However, when it comes to the slow-moving weak target detection, MTI will suppress not only the strong echoes from the static IOs but also the weak echoes from the TOIs, leading to the loss in the detection probability. Therefore, a more reliable clutter suppression method is needed. \par
In recent years, a number of low-rank and sparse matrix decomposition (LRaSMD) optimization algorithms have been proposed, which concentrate on developing fast approximations and meaningful decompositions \cite{Zhang2016lrasmd}. Go decomposition (Godec) is a robust LRaSMD method that can be used for matrix completion, background modeling, shadow/light removal, etc. \cite{zhou2011godec}. Naive Godec has a high computational complexity due to the singular value decomposition (SVD), and the authors in \cite{zhou2011godec} significantly reduced its time cost by replacing SVD with bilateral random projection (BRP). Besides, the power scheme modification is also adopted to achieve faster convergence of Godec, which also contributes to reducing the time cost to some extent. \par
When it comes to the application of Godec in radar detection, the authors in \cite{Su2021rpca} applied Godec to the moving targets radar detection and compared the results with those obtained by the MTI-based method, demonstrating the superior performance of the proposed Godec-based method for detecting the slow-moving targets. However, the time cost of the Godec-based method, which is also an important consideration in practical applications, was not mentioned. In addition, the authors in \cite{Chang2021godec} noted that finding appropriate values for the parameters of Godec is very challenging and Godec does not provide any guideline for determining those values, which to some extent limits the application of Godec in practical radar detection. Therefore, a method for obtaining the relevant parameters of Godec needs to be clarified to enhance the practicality of the Godec-based method. \par
In this paper, we utilized a wideband LFMCW radar for simulation and a pulsed radar for experiment validation. The main contributions can be summarized as follows:
\begin{itemize}
\item[$\bullet$] A Godec-based clutter suppression solution is proposed for the slow-moving weak target detection. New iteration termination conditions are incorporated into the original Godec algorithm, enabling its application to slow-moving weak target radar detection with the presence of statistical clutter. Besides, the time complexity is discussed. Furthermore, we show that the given solution can be combined with the efficient cell-averaging constant false alarm rate (CA-CFAR) for the slow-moving weak target detection. \par
\item[$\bullet$] To address the challenge of selecting the sparsity upper bound constraint in Godec, we employ a piecewise approach with an auxiliary parameter to reduce the times of traversal in the searching process. Additionally, a normalized performance function is introduced to determine the optimal value of the auxiliary parameter, thereby achieving the best CFAR detection results. \par
\item[$\bullet$] Simulation results are presented to show the reliability of the proposed solution compared to the traditional ones based on MTI. Additionally, tradeoffs among the number of the false alarm cells, the detection performance and the time consumption are revealed, guiding parameter settings of the proposed solution. Moreover, experiment validation is conducted, which not only verifies the effectiveness of the proposed solution in practical applications, but also provides further insights into the scenarios where the proposed solution is most applicable. \par
\end{itemize}
\par
The remainder of this paper is organized as follows. Section II details the relevant preliminaries. Section III outlines the principle of Godec and describes the proposed Godec-based clutter suppression scheme, highlighting the novel parameters introduced into the original Godec algorithm and the methodology employed for searching the optimal auxiliary parameter. Section IV and V respectively present the simulation results and the experiment validation, along with a detailed analysis. Finally, Section VI concludes the paper. \par

\section{Preliminary}
\subsection{System Settings}
\begin{figure}[htbp]
\centerline{\includegraphics[width=0.45\textwidth]{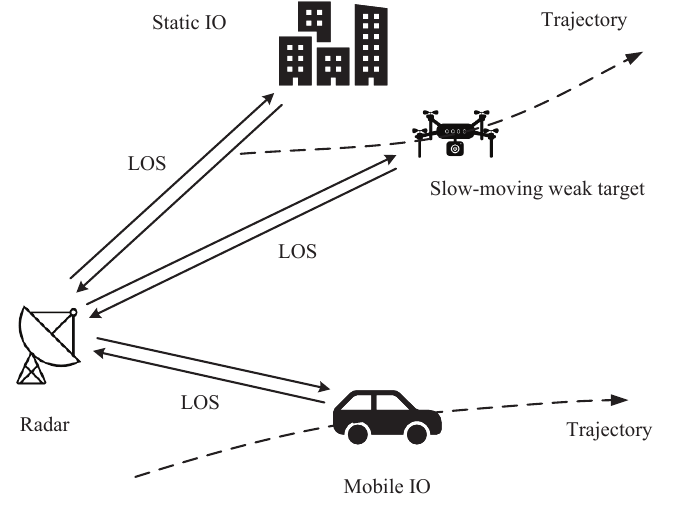}}
\caption{Diagram of the slow-moving weak target detection.}
\label{system_model}
\end{figure}

As shown in Fig. \ref{system_model}, there exists at least one radar, one static IO and one slow-moving weak target, and the mobile IOs may also exist in the detection area. We define the RCS of the slow-moving weak target as no more than 0 dBsm, with a velocity not exceeding 1 m/s. Objects that do not meet these criteria are all considered IOs \cite{Yang2019rcs,Pieraccini2017rcs}. To show the masking effects, the range between the slow-moving weak target and the radar is set close to that between the static IO and the radar. Here we assume that the difference between these two ranges is not above 1 m. The radar can both transmit and receive signals, and only the line-of-sight (LOS) signals are considered. \par
\subsection{Signal Processing}
\begin{figure}[htbp]
\centerline{\includegraphics[width=0.75\textwidth]{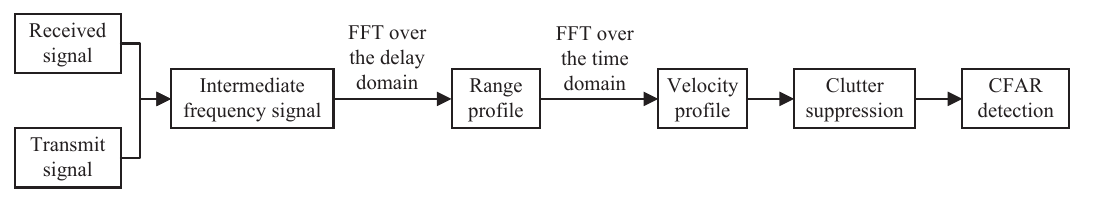}}
\caption{The general structure of radar signal processing.}
\label{lfmcw_processing}
\end{figure}

In this part, we mainly discuss the traditional signal processing scheme for LFMCW radar in the slow-moving weak target detection. The general structure of signal processing is shown in Fig. \ref{lfmcw_processing}. \par
\subsubsection{Weibull Clutter}
\begin{figure}[htbp]
\centerline{\includegraphics[width=0.75\textwidth]{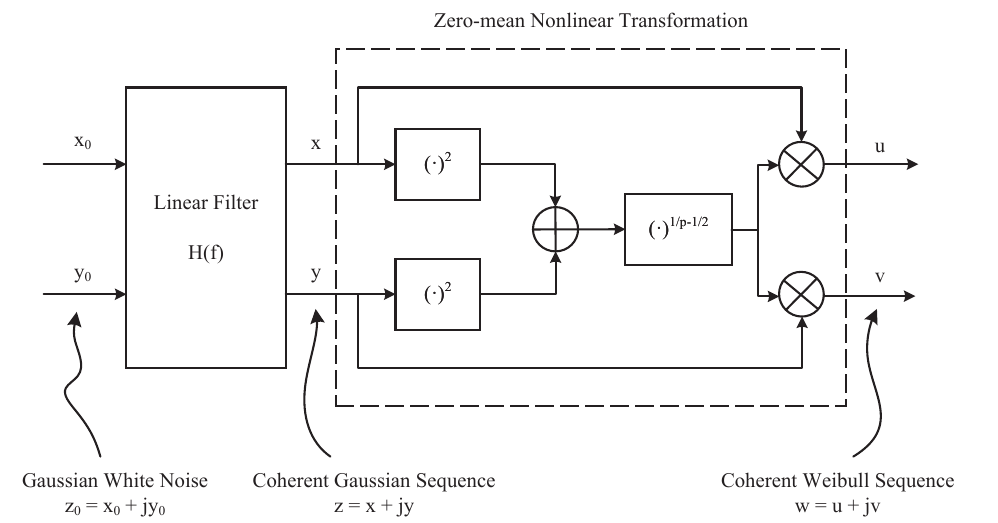}}
\caption{Diagram of the coherent Weibull clutter generation.}
\label{Weibull_generation}
\end{figure}

Unlike most radar target detection researches that only consider the background noise, this paper also takes into account the statistical clutter received from the environment, specifically the clutter with an amplitude that follows a Weibull distribution. Here we consider the generation of the coherent Weibull clutter using the zero-mean nonlinear transformation (ZMNL) method \cite{Huang2024weibull}, and the block diagram is shown in Fig. \ref{Weibull_generation}. The real and imaginary part of the coherent Weibull sequence $u$ and $v$ can be expressed as
\begin{equation}
\begin{cases}
u = x(x^2+y^2)^{1/p-1/2} \\
v = y(x^2+y^2)^{1/p-1/2}
\end{cases}
\label{Weibull_expression}
\end{equation}
where $x$ and $y$ are zero mean joint Gaussian sequences, $p$ represents the shape parameter of Weibull distribution. \par
As illustrated in Fig. \ref{Weibull_generation}, prior to applying the ZMNL, two independent Gaussian white noise channels are firstly processed through a linear filter $H(f)$. This step determines the power spectrum characteristics of the Weibull clutter, including Gaussian, exponential, and all-pole, etc. \cite{Melebari2015psdclutter}. The power spectral density (PSD) functions for these three types of spectra are presented below:
\begin{equation}
\text{Gaussian:} \qquad S_\text{g}(f) = H_\text{g}^2(f) = C_0\exp(\frac{f^2}{2f_{\text{3dB}}^2})
\label{Gaussian_psd}
\end{equation}
\begin{equation}
\text{Exponential:} \qquad S_\text{e}(f) = H_\text{e}^2(f) = C_0\exp(-\frac{\vert f \vert}{f_{\text{3dB}}})
\label{Exponential_psd}
\end{equation}
\begin{equation}
\text{All-pole:} \qquad S_\text{a}(f) = H_\text{a}^2(f) = \frac{C_0}{1+\vert \frac{f}{f_{\text{3dB}}} \vert^\gamma}
\label{All_pole_psd}
\end{equation}
where $\gamma$ is the exponent of the all-pole model, $C_0$ is the average clutter power, which is expressed as \cite{weibullparasetting2012}
\begin{equation}
C_0 = \frac{w_\text{m}^2\Gamma(1+2/p)}{(\ln 2)^{2/p}}
\label{average_power}
\end{equation}
where $w_{\text{m}}$ represents the median amplitude of Weibull clutter, which is expressed as
\begin{equation}
w_{\text{m}} = q(\ln 2)^{1/p}
\label{median_Weibull}
\end{equation}
where $q$ represents the scale parameter of Weibull distribution. $f_{\text{3dB}}$ represents the spectral width corresponding to the half-power point of the clutter spectrum. As a rule of thumb, it is approximately $3\%$ of the Doppler frequency of wind \cite{Zhang2008vw}, which can be expressed as
\begin{equation}
f_{\text{3dB}} = \frac{2v_{\text{w}}}{\lambda}*0.03
\label{3dB_parameter}
\end{equation}
where $v_{\text{w}}$ represents the relative radial velocity of wind on average, which is the main cause of the Weibull clutter considered in this paper. $\lambda$ is the carrier wavelength. \par

\subsubsection{Range-velocity Map}
The transmit signal in the $m$-th chirp of LFMCW radar $s_{\text{t}}(t)$ is expressed as
\begin{equation}
s_{\text{t}}(t) = A_{\text{t}}\exp\big(j2\pi(f_0t + \mu t^2/2)\big), (m-1)T_{\text{c}}\leq t \leq mT_{\text{c}}
\label{tx_signal}
\end{equation}
where $A_{\text{t}}$ is the signal amplitude, $T_{\text{c}}$ is the duration of each chirp, $f_0$ is the carrier frequency, $\mu = B/T_{\text{c}}$ is the linear modulation slope ($B$ is the signal bandwidth).
Assuming that there are $N_{\text{IO}}$ IOs in the detection area, the received signal $r(t)$ can be expressed as
\begin{equation}
r(t) = \underbrace{A_0s_{\text{t}}\big(t-\tau_0(t)\big)}_{\text{TOI}} + \underbrace{\sum\limits_{i=1}^{N_{\text{IO}}}A_is_{\text{t}}\big(t-\tau_i(t)\big)}_{\text{IOs}} + c(t) + w(t)
\label{received_signal}
\end{equation}
where $c(t)$ is the Weibull clutter, $w(t)$ is the background noise, $A_0, A_1, \ldots, A_{N_{\text{IO}}}$ are the amplitudes of received signals, which can be obtained by the radar equation \cite{skolnik1970radar}, $\tau_0, \tau_1, \ldots, \tau_{N_{\text{IO}}}$ are the signal propagation delays, which can be expressed as
\begin{equation}
\tau_i(t) = 2(R_i + v_it)/c
\label{time_delay}
\end{equation}
where $c$ is the light speed, $R_i$ and $v_i$ are respectively the range and relative radial velocity between the radar and the $i$-th object. \par
After receiving the signals reflected from the objects, a mixture between the received signal $r(t)$ and the transmit signal $s_{\text{t}}(t)$ is conducted, and the intermediate frequency (IF) signal $r_{\text{IF}}(t)$ can be obtained, which is expressed as
\begin{equation}
r_{\text{IF}}(t) = r(t)s_{\text{t}}^*(t) = \underbrace{A_{\text{IF}_0}s_{\text{IF}_0}(t)}_{\text{TOI}} + \underbrace{\sum\limits_{i=1}^{N_{\text{IO}}}A_{\text{IF}_i}s_{\text{IF}_i}(t)}_{\text{IOs}} + \tilde c(t) + \tilde w(t)
\label{received_IF_signal}
\end{equation}
where $\tilde c(t)$ and $\tilde w(t)$ are the Weibull clutter and noise, $A_{\text{IF}_0}, A_{\text{IF}_1}, \ldots, A_{\text{IF}_{N_{\text{IO}}}}$ are the amplitudes of the IF signals $s_{\text{IF}_0}(t), s_{\text{IF}_1}(t), \ldots, s_{\text{IF}_{N_{\text{IO}}}}(t)$, and the IF signals are expressed as
\begin{equation}
s_{\text{IF}_i}(t) = \exp\Big(-j2\pi\big(f_0\tau_i(t) + \mu\tau_i(t)t - \mu\tau_i^2(t)/2\big)\Big), i = 0,1,\ldots,N_{\text{IO}}
\label{separated_IF_signal}
\end{equation}
\par
By sampling $r_{\text{IF}}(t)$ and reshaping it into an $M \times N$ matrix, the two-dimensional discrete signal matrix $\mathbf{R}_{\text{IF}}(m,n)$ can be obtained, which is expressed as
\begin{equation}
\mathbf{R}_{\text{IF}}(m,n) = r_{\text{IF}}\big((m-1)T_{\text{c}}/M + (n-1)T_{\text{c}}\big), m = 1,2,\ldots,M; n = 1,2,\ldots,N
\label{2D_discrete_matrix}
\end{equation}
where $M$ and $N$ respectively represent the number of samples per chirp and the number of sampling chirps per radar scan. By performing the two-dimensional FFT after adopting the Hamming window on $\mathbf{R}_{\text{IF}}(m,n)$, the range and velocity profiles can be obtained, i.e., the range-velocity map. \par
\subsubsection{MTI}
For the slow-moving weak target detection, as the velocity of the target is close to zero, and the static IOs within the detection area are strong reflectors, we usually need to suppress the strong echoes to guarantee the correct detection of the weak target. MTI is a common method for suppressing the static clutter, which utilizes the difference in Doppler frequencies between the echoes of the static and moving objects \cite{Mao2023mti}. By designing an finite impulse response filter in the Doppler domain, the frequency response of the filter has a deep stop-band near the zero frequency, thereby suppressing the echoes of the static IOs. \par
\begin{figure}[htbp]
\centerline{\includegraphics[width=0.5\textwidth]{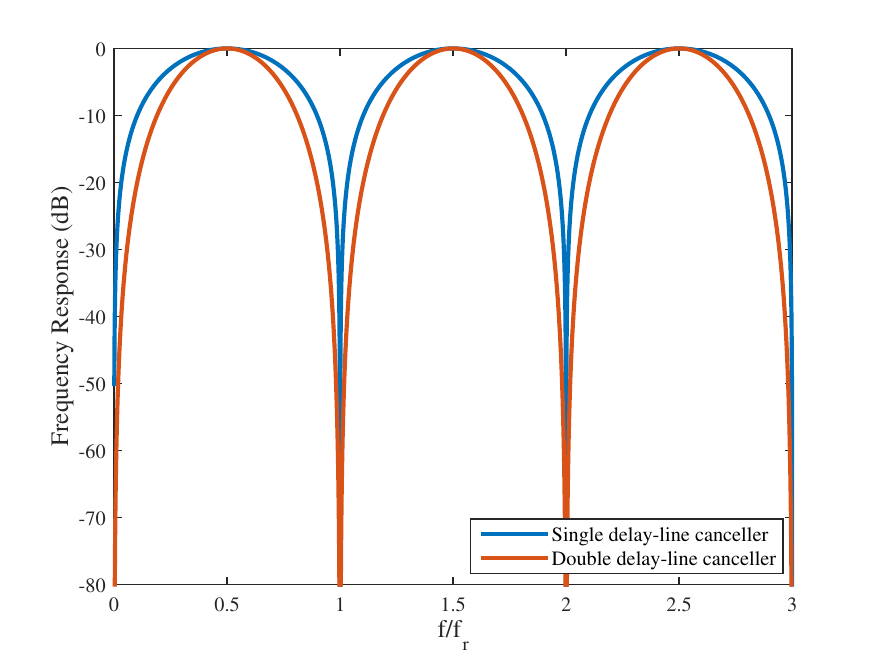}}
\caption{Frequency responses of the common MTI filters.}
\label{mti_filters}
\end{figure}

Many MTI filters have been designed, among which the simplest and most commonly used ones are the single and double delay-line canceller \cite{Ghorbani2020mti}. The frequency responses of these two cancellers are shown in Fig. \ref{mti_filters}, where $f_{\text{r}}$ represents the chirp repetition frequency. It can be seen that the more chirps involved in the canceller, the deeper the stop-band nearby the zero frequency, which is detrimental to detecting the slow-moving weak target. Therefore, only the single delay-line canceller is considered in this article. \par
When processing an $M \times N$ range-velocity map, MTI performs a difference operation between every two adjacent columns of the matrix. Therefore, the number of floating-point operations (flops) required by MTI to process a single range-velocity map is $(N-1)M$.
\subsubsection{CFAR}
\begin{figure}[htbp]
\centerline{\includegraphics[width=0.5\textwidth]{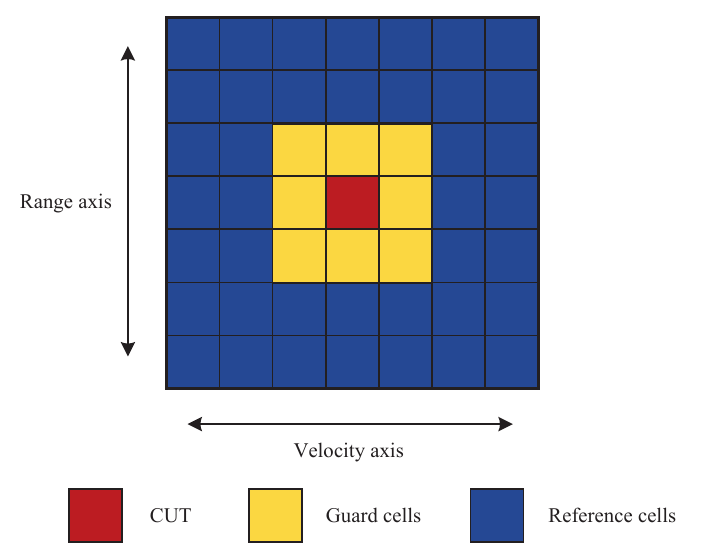}}
\caption{2-D detection window of CA-CFAR and OS-CFAR.}
\label{CFAR_window}
\end{figure}

As shown in Fig. \ref{lfmcw_processing}, the CFAR detection is performed after the clutter suppression. Two kinds of CFAR are considered including the CA-CFAR and the ordered-statistic CFAR (OS-CFAR) \cite{Zhang2019cacfar,Safa2021oscfar}. The 2-D detection window of the two CFAR detectors are shown in Fig. \ref{CFAR_window}, where the presence of TOI in the cell under test (CUT) is determined by comparing its intensity with the average intensity of the reference cells (CA-CFAR) / the intensity of the $k$-th strongest reference cell (OS-CFAR). It has been revealed that the OS-CFAR achieves better performance against the masking effect but higher computational complexity than the CA-CFAR \cite{Rohling2013maskingeffect,Jeong2022complexity}. \par
When processing an $M \times N$ range-velocity map, CA-CFAR requires $N_{\text{R}}N_0$ flops to compute the average intensity of all the $N_0$ detection windows, where each detection window contains $N_{\text{R}}$ reference cells. OS-CFAR requires sorting the intensities of all $N_{\text{R}}$ reference cells within each detection window, resulting in $\mathcal{O}(N_{\text{R}}\log_2(N_{\text{R}})) \times N_0$ comparisons. \par
\section{Proposed Godec-based Clutter Suppression}
In this section, a reliable slow-moving weak target detection scheme based on the Godec clutter suppression and CA-CFAR is proposed.
\subsection{Principle of Godec}
For the problem of a matrix $\mathbf{X}$ which can be expressed as
\begin{equation}
\mathbf{X} = \mathbf{L} + \mathbf{S} + \mathbf{N}
\label{godec_model}
\end{equation}
where $\mathbf{L}$, $\mathbf{S}$ and $\mathbf{N}$ are respectively the low-rank, sparse and noise part of the matrix, Godec can efficiently and robustly estimate the low-rank part $\mathbf{L}$ and the sparse part $\mathbf{S}$ by alternatively assigning the low-rank approximation of $\mathbf{X} - \mathbf{S}$ to $\mathbf{L}$ and the sparse approximation of $\mathbf{X} - \mathbf{L}$ to $\mathbf{S}$ \cite{zhou2011godec}. The LRaSMD problem can be solved by an optimization, expressed as
\begin{equation}
\begin{aligned}
\min_{\mathbf{L}, \mathbf{S}}\quad & \Vert\mathbf{X} - \mathbf{L} - \mathbf{S}\Vert_{\text{F}}^2 \\
\text{s.t.} \quad & \text{rank}\big(\mathbf{L}\big) \leq r \\
  & \text{card}\big(\mathbf{S}\big) \leq s \\
\end{aligned}
\end{equation}
where the notation $\Vert\cdot\Vert_{\text{F}}$ represents the Frobenius norm of its argument, rank($\cdot$) and card($\cdot$) respectively represent the rank and cardinality of the matrix, $r$ and $s$ respectively represent the rank upper bound constraint on $\mathbf{L}$ and the cardinality upper bound constraint on $\mathbf{S}$. \par
BRP and the power scheme modification are adopted to solve this optimization problem efficiently, and the details of the BRP-based Godec algorithm can be found in \cite{zhou2011godec}. Here we define $\zeta$ $(\zeta \geq 0)$ as the integer exponent of the power scheme modification. Besides, the decomposition error of Godec $\epsilon$ is defined as
\begin{equation}
\epsilon = \Vert \mathbf{X} - \mathbf{L} - \mathbf{S} \Vert_{\text{F}}/\Vert \mathbf{X} \Vert_{\text{F}} \times 100\%
\end{equation}
\subsection{Applications of Godec for Clutter Suppression}
As mentioned in \cite{Su2021rpca}, due to the low-rank and sparse structure of the static and moving objects in range-velocity maps across multiple radar scans, Godec is suitable for suppressing the echoes of the static objects. \par
\subsubsection{Procedures}
\begin{figure}[htbp]
\centerline{\includegraphics[width=1.0\textwidth]{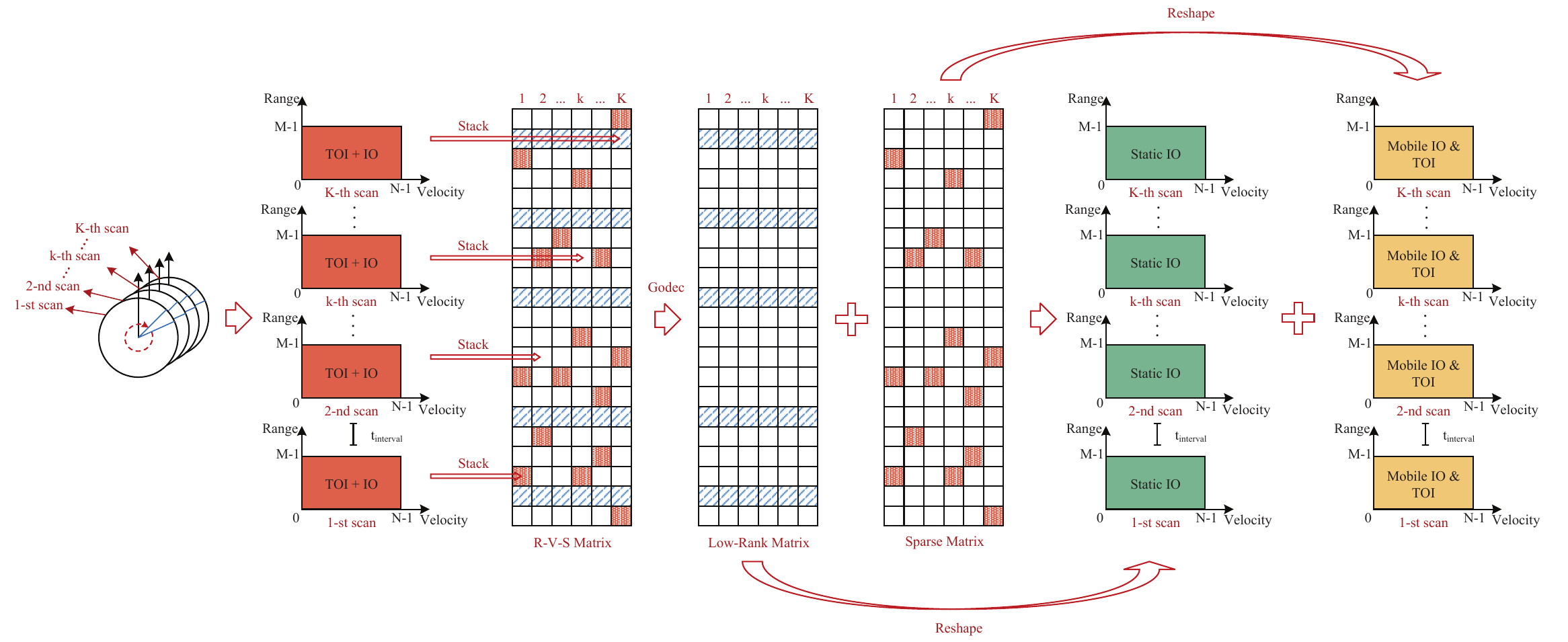}}
\caption{Diagram of the Godec-based clutter suppression.}
\label{godec_diagram}
\end{figure}

As shown in Fig. \ref{godec_diagram}, we obtain $K$ range-velocity maps through $K$ radar scans, with a time interval $t_{\text{interval}}$ between each two scans. To fully explore the capability of Godec in suppressing the stationary clutter, the TOI is required to move across at least one range bin between two adjacent scans. By stacking each range-velocity map into a column vector and combining $K$ column vectors into a ``range-velocity-scan'' matrix $\mathbf{X}_{\text{rvs}}$, Godec can be performed on $\mathbf{X}_{\text{rvs}}$ to obtain the low-rank part $\mathbf{L}_{\text{rvs}}$ and the sparse part $\mathbf{S}_{\text{rvs}}$. Then, by inversely reshaping $\mathbf{L}_{\text{rvs}}$ and $\mathbf{S}_{\text{rvs}}$, the static and dynamic parts of the range-velocity maps can be obtained, corresponding to containing the static and mobile objects, respectively. \par

\subsubsection{Parameters}
Since we only perform CFAR detection on the obtained dynamic parts of the range-velocity maps, and the value of the sparsity upper bound constraint $s$ directly affects the ability of Godec to decompose the dynamic parts, this paper focuses solely on investigating the searching method for the parameter $s$. Other related parameters, such as $r$, $\zeta$, etc., are set to fixed values. To facilitate the rational setting of $s$ and improve the efficiency of the proposed Godec-based solution in reliable target detection, the following parameters are added into the original Godec algorithm:
\begin{itemize}
\item[$\bullet$] $iter_{\text{max}}$, $iter_{\text{cov}}$, $\delta$:
Due to the presence of Weibull clutter, the stopping criterion for iteration of Godec in \cite{zhou2011godec} is usually not achievable. Therefore, to explore the optimal efficiency of the slow-moving weak target detection based on the Godec clutter suppression, three new parameters $iter_{\text{max}}$, $iter_{\text{cov}}$ and $\delta$, which respectively represent the maximum iteration times, the iteration times required for convergence and the threshold for the error difference between two adjacent iterations, are added into the original Godec algorithm. Accordingly, two additional stopping criteria are introduced in the original Godec algorithm: one is when the number of iterations exceeds $iter_{\text{max}}$, and the other is when the difference in decomposition error between two consecutive iterations is smaller than $\delta$, and $iter_\text{cov}$ will be output when the iteration terminates. \par
\item[$\bullet$] $N_{\text{mov}}$:
Due to the large number of cells contained in $\mathbf{X}_{\text{rvs}}$, and the fact that one strong mobile IO often corresponds to multiple dynamic cells due to its significant spectral leakage during the two-dimensional FFT, using an exhaustive searching method to determine the value of the sparsity upper bound constraint $s$ is extremely computationally expensive. Accordingly, we employ a piecewise approach to search the reasonable value for $s$ in order to reduce the times of traversal, and the obtained value for $s$ exhibit a certain degree of robustness to the dynamic statistical clutter in the same scenario. \par
Considering the characteristic that the echoes of strong mobile IOs exhibit significant spectral leakage in the velocity domain, we define a new integer parameter $N_{\text{mov}}$ $(N_{\text{mov}} \in (0,M])$ that satisfies the following relationship with $s$:
\begin{equation}
s = N_{\text{mov}} \times N \times K
\label{N_mov}
\end{equation}
In (\ref{N_mov}), we assume that the horizontal bands caused by the spectral leakage stretch over the entire velocity domain, which is the worst situation. Compared to the exhausting traversal of $s$, the traversal of the integer $N_\text{mov}$ is much more efficient. \par
\end{itemize}
\subsubsection{Algorithms}
Based on the parameters above, the Godec algorithm adapted for slow-moving weak target detection is presented in \textbf{Algorithm~\ref{godec_details}}, where $\mathbf{A}_1 \in \mathbb{R}^{K \times r}$ and $\mathbf{A}_2 \in \mathbb{R}^{MN \times r}$ mentioned at line 3 and line 6 are Gaussian random matrices in BRP. The QR decomposition mentioned at line 7 is adopted to extract the rank-$r$ components from $\mathbf{X}_{\text{rvs}}$, and the approximate low-rank matrix is recovered at line 8. At line 9, after eliminating the approximate low-rank part, the remaining components of $\mathbf{X}_\text{rvs}$ are sorted by intensity, and the top $ (N_\text{mov} \times N \times K)$ strongest cells are extracted as the components of the sparse matrix. \par
\begin{algorithm}[htbp]
    \textbf{Input:} ~~
                    \begin{itemize}
                        \setlength{\itemsep}{0pt} 
                        \setlength{\parsep}{0pt}
                        \setlength{\leftmargin}{2em} 
                        \renewcommand{\labelitemi}{} 
                        \item $N$: the number of sampling chirps per radar scan;
                        \item $K$: the number of radar scans;
                        \item $r$: the rank upper bound constraint on the low-rank matrix $\mathbf{L}_{\text{rvs}}$;
                        \item $N_{\text{mov}}$: the integer parameter correlated with the sparsity upper bound constraint $s$;
                        \item $\zeta$: the integer exponent of the power scheme modification;
                        \item $\epsilon_0$: the pre-determined decomposition error bound;
                        \item $iter_{\text{max}}$: the maximum iteration times;
                        \item $\delta$: the threshold for the error difference between two adjacent iterations;
                        \item $\mathbf{X}_{\text{rvs}}$: the ``range-velocity-scan'' matrix
                    \end{itemize}
	\textbf{Output:} ~~
                    \begin{itemize}
                        \setlength{\itemsep}{0pt} 
                        \setlength{\parsep}{0pt}
                        \setlength{\leftmargin}{2em} 
                        \renewcommand{\labelitemi}{} 
                        \item $iter_{\text{cov}}$: the iteration times required for convergence;
                        \item $\mathbf{L}_{\text{rvs}}$: the low-rank matrix;
                        \item $\mathbf{S}_{\text{rvs}}$: the sparse matrix;
                        \item $\epsilon$: the decomposition error
                    \end{itemize}
	\caption{Godec algorithm adapted for slow-moving weak target detection}
    \label{godec_details}
    \vspace{-7pt}
    \rule{\linewidth}{0.4pt} 
    \vspace{-12pt}
    \begin{algorithmic}[1] 

        \STATE \textbf{Initialize} $\mathbf{L}_0 = \mathbf{X}_{\text{rvs}}$, $\mathbf{S}_0 = \mathbf{0}$, $t = 0$;

        \WHILE {$( \Vert \mathbf{X}_{\text{rvs}} - \mathbf{L}_t - \mathbf{S}_t \Vert_{\text{F}}/\Vert \mathbf{X}_{\text{rvs}} \Vert_{\text{F}} \geq \epsilon_0) \wedge (t \leq iter_\text{max}) \wedge (\Big\vert \frac{\Vert \mathbf{X}_{\text{rvs}} - \mathbf{L}_{t-1} - \mathbf{S}_{t-1} \Vert_{\text{F}} - \Vert \mathbf{X}_{\text{rvs}} - \mathbf{L}_{t} - \mathbf{S}_{t} \Vert_{\text{F}}}{\Vert \mathbf{X}_{\text{rvs}} \Vert_{\text{F}}}\Big\vert \geq \delta) $}
            \STATE Generate a Gaussian random matrix $\mathbf{A}_1 \in \mathbb{R}^{K \times r}$;
            \STATE $t = t + 1$;
            \STATE $\tilde{\mathbf{L}} = \Big[\big(\mathbf{X}_{\text{rvs}} - \mathbf{S}_{t-1}\big)\big(\mathbf{X}_{\text{rvs}} - \mathbf{S}_{t-1}\big)^{\text{T}}\Big]^{\zeta}\big(\mathbf{X}_{\text{rvs}} - \mathbf{S}_{t-1}\big)$;
            \STATE $\mathbf{Y}_1 = \tilde{\mathbf{L}}\mathbf{A}_1$, $\mathbf{A}_2 = \mathbf{Y}_1$;
            \STATE $\mathbf{Y}_2 = \tilde{\mathbf{L}}^{\text{T}}\mathbf{Y}_1 = \mathbf{Q}_2\mathbf{R}_2$, $\mathbf{Y}_1 = \tilde{\mathbf{L}}\mathbf{Y}_2 = \mathbf{Q}_1\mathbf{R}_1$;

            \STATE $\mathbf{L}_t = \mathbf{Q}_1\Big[\mathbf{R}_1\big(\mathbf{A}_2^\text{T}\mathbf{Y}_1\big)^{-1}\mathbf{R}_2^{\text{T}}\Big]^{1/(2\zeta+1)}\mathbf{Q}_2^{\text{T}}$;
            \STATE $\mathbf{S}_t = \mathcal{P}_\Omega\big(\mathbf{X}_{\text{rvs}}-\mathbf{L}_t\big)$, $\Omega$ is the nonzero subset of the first $(N_\text{mov} \times N \times K)$ largest entries of $\vert \mathbf{X}_{\text{rvs}}-\mathbf{L}_t \vert$;
        \ENDWHILE

        \STATE $iter_{\text{cov}} = t-1$, $\mathbf{L}_{\text{rvs}} = \mathbf{L}_{t}$, $ \mathbf{S}_{\text{rvs}} = \mathbf{S}_{t}$;
        \STATE $\epsilon = \Vert \mathbf{X}_{\text{rvs}} - \mathbf{L}_{\text{rvs}} - \mathbf{S}_{\text{rvs}} \Vert_{\text{F}}/\Vert \mathbf{X}_{\text{rvs}} \Vert_{\text{F}} \times 100\%$.
    \end{algorithmic}
\end{algorithm}
As mentioned in \cite{zhou2011godec}, when applying the power scheme modification ($\zeta > 0$), the Godec algorithm presented in \textbf{Algorithm~\ref{godec_details}} requires $[r^2(MN+3K+4r)+4(\zeta+1)MNKr](iter_{\text{cov}}+1)$ flops, where $iter_\text{cov}$ has a significant impact on the time complexity of the Godec-based solution. Besides, there is an intrinsic tradeoff between the false alarm rate $P_\text{fa}$ and the detection probability $P_{\text{d}}$ in CFAR detection. In the proposed Godec-based solution, all these three factors ($iter_\text{cov}$, $P_\text{fa}$ and $P_{\text{d}}$) are associated with the value of $s$. Therefore, searching for the optimal value of $s$ (or $N_\text{mov}$) often requires balancing these three factors, with the goal of maximizing the detection probability of TOI while minimizing the values of the other two factors as much as possible, thereby ensuring both the reliability and efficiency of the algorithm. \par
We use the average number of false alarm cells in all scans $N_\text{fa}$ as the factor for measuring false alarms instead of $P_\text{fa}$ in this paper. Considering the factors related to $N_{\text{mov}}$ mentioned above, we define a normalized performance function $f$ here to quantify the performance of the proposed Godec-based solution for different values of $N_{\text{mov}}$, which is expressed as
\begin{equation}
f = 0.5*\Big(1/iter_\text{cov} + 0.5*\big(P_\text{d} + 1/(N_\text{fa} + 1)\big)\Big)
\label{performance_func}
\end{equation}
where $iter_\text{cov} \in [1,iter_\text{max}]$, $P_\text{d} \in [0,1]$ and $N_\text{fa} \geq 0$. The aim of searching for the optimal $N_\text{mov}$ is to maximize $f$. 
Accordingly, the details of the slow-moving weak target detection based on Godec clutter suppression and CA-CFAR can be found in \textbf{Algorithm~\ref{godec_scheme}}, in which the process of searching for the optimal parameter $N_{\text{mov}}$ is from line 1 to line 13. \par
\begin{algorithm}[htbp]
    \textbf{Input:} ~~
                    \begin{itemize}
                        \setlength{\itemsep}{0pt} 
                        \setlength{\parsep}{0pt}
                        \setlength{\leftmargin}{2em} 
                        \renewcommand{\labelitemi}{} 
                        \item $M$: the number of samples per chirp;
                        \item $N_{\text{G}}$: the number of guard cells;
                        \item $N_{\text{R}}$: the number of reference cells;
                        \item $P_{\text{fa}}$: the false alarm rate;
                        \item Parameters input in \textbf{Algorithm~\ref{godec_details}}
                    \end{itemize}
	\textbf{Output:} ~~
                    \begin{itemize}
                        \setlength{\itemsep}{0pt} 
                        \setlength{\parsep}{0pt}
                        \setlength{\leftmargin}{2em} 
                        \renewcommand{\labelitemi}{} 
                        \item $\mathbf{X}_{0}$: the CFAR detection result with the best performance
                    \end{itemize}
	\caption{Slow-moving weak target detection based on Godec clutter suppression and CA-CFAR}
    \label{godec_scheme}
    \vspace{-7pt}
    \rule{\linewidth}{0.4pt} 
    \vspace{-12pt}
    \begin{algorithmic}[1] 

        \STATE \textbf{Initialize} $N_{\text{mov}_0} = M-1$ and $f_0 = 0$;

        \WHILE {$N_{\text{mov}} > 0$}
            \STATE Perform \textbf{Algorithm~\ref{godec_details}} and obtain $iter_\text{cov}$;
            \STATE Inversely reshape $\mathbf{L}_{\text{rvs}}$ and $\mathbf{S}_{\text{rvs}}$ to get $\mathbf{L}_{\text{rv}}$ and $\mathbf{S}_{\text{rv}}$ in $K$ radar scans;
            \STATE $\mathbf{X}_{\text{det}} = \text{CA-CFAR}\big(\mathbf{S}_{\text{rv}}, N_\text{G}, N_\text{R}, P_{\text{fa}}\big)$;
            \STATE Calculate the average false alarm cells in $K$ scans $N_\text{fa}$ and the detection probability of TOI $P_\text{d}$;
            \STATE Calculate $f$ according to (\ref{performance_func});

            \IF {$f > f_0$}
                \STATE $f_0 = f$;
                \STATE $N_{\text{mov}_0} = N_{\text{mov}}$;
            \ENDIF
                \STATE $N_{\text{mov}} = N_{\text{mov}} -1 $;
        \ENDWHILE

        \STATE Perform \textbf{Algorithm~\ref{godec_details}} with $N_{\text{mov}_0}$;
        \STATE Inversely reshape $\mathbf{L}_{\text{rvs}}$ and $\mathbf{S}_{\text{rvs}}$ to get $\mathbf{L}_{\text{rv}}$ and $\mathbf{S}_{\text{rv}}$ in $K$ scans;
        \STATE $\mathbf{X}_{0} = \text{CA-CFAR}\big(\mathbf{S}_{\text{rv}}, N_\text{G}, N_\text{R}, P_{\text{fa}}\big)$.
    \end{algorithmic}
\end{algorithm}
\section{Simulation Results}
\subsection{Simulation Settings}
\begin{figure}[htbp]
\centering
\centerline{\includegraphics[width=0.5\textwidth]{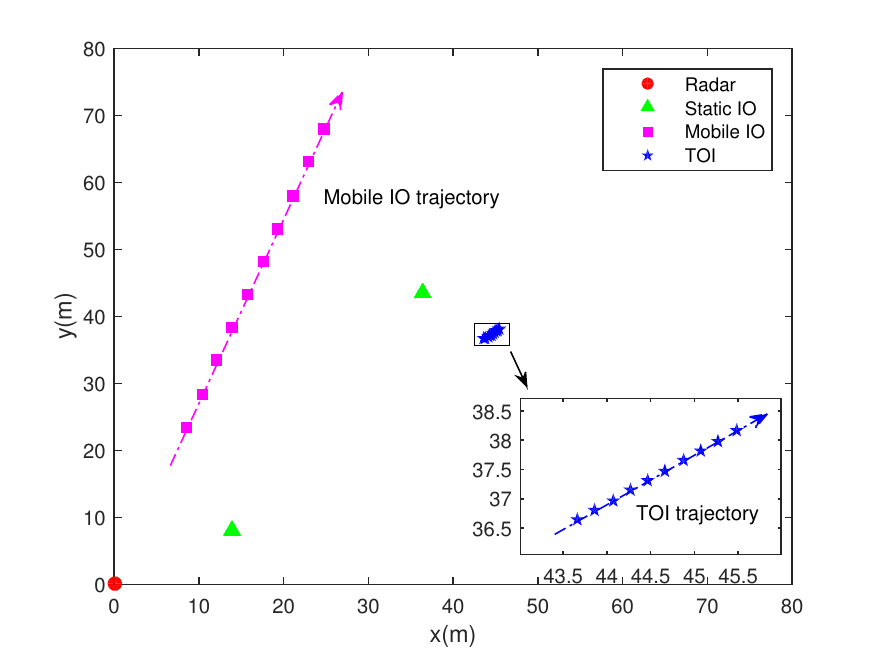}}
\caption{Radar detection network simulation in 10 scans.}
\label{Network_simulation}
\end{figure}

In this section, we show the simulation results in typical target detection schemes and the proposed Godec-based scheme, and demonstrate the performance advantages of the proposed one. As shown in Fig. \ref{Network_simulation}, at the size of $80\text{m}\times80\text{m}$, the simulation network has 1 radar located at the original point. In the detection area, there are two static IOs, one mobile IO and one TOI. To simplify the problem, we assume that the TOI and the mobile IO are respectively moving away from the radar at a constant speed. The target and the mobile IO trajectory formed in 10 radar scans can be seen in Fig. \ref{Network_simulation} as the dash-dotted line. The radar and object simulation parameters are detailed in Table \ref{Radar_parameters} and Table \ref{Object_parameters}, respectively. \par
\begin{figure}[htbp]
	\centering
	\subfloat[PDF]{
	\includegraphics[width=0.5\textwidth]{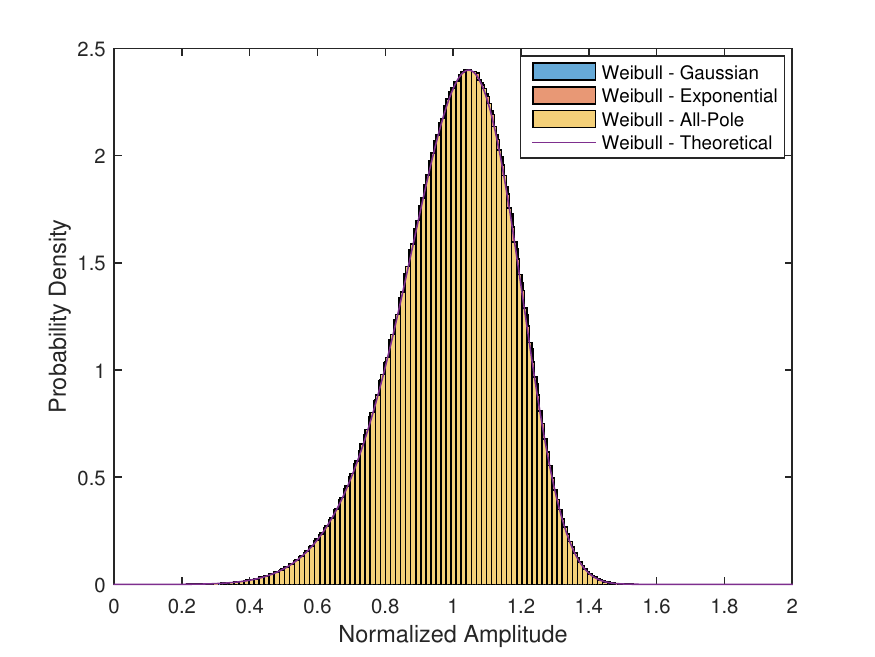}
    \label{PDF}}
    \subfloat[PSD]{
	\includegraphics[width=0.5\textwidth]{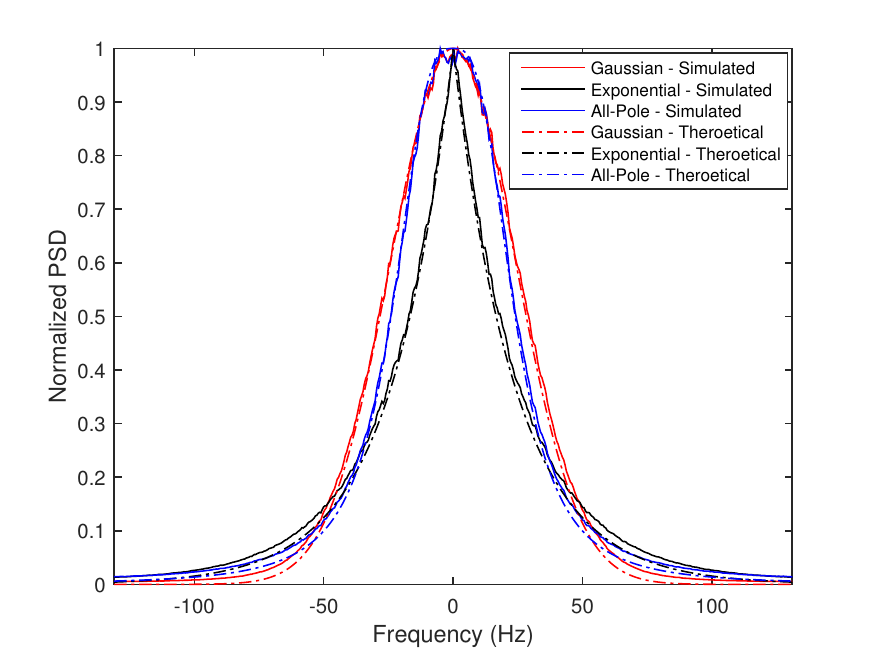}
    \label{PSD}}
   \quad  
	\caption{Weibull clutter ($p$ = 1.6, $q$ = 6.9) simulation verification.}
	\label{Weibull_simulation_verification}
\end{figure}

\begin{table}[htbp]
\caption{Radar Simulation Settings}
\centering
\begin{tabular}{lll}
\toprule
Parameters & Description & Value \\
\midrule
$P_\text{t}$ & Radar transmit power (dBm) & 20 \\
$G_\text{t}$ & Transmit antenna gain (dB) & 20 \\
$G_\text{r}$ & Receive antenna gain (dB) & 20 \\
$N_{0}$ & Power spectrum density of noise (dBm/Hz) & -174 \\
$BW$ & Bandwidth (GHz) & 1 \\
$f_{\text{c}}$ & Carrier frequency (GHz) & 24 \\
$M$ & Sampling ADC number per chirp & 1024 \\
$N$ & Sampling chirp number per scan & 256 \\
$K$ & Number of radar scans & 10 \\
$T_{\text{r}}$ & Chirp repetition interval (ms) & 0.1 \\
$t_{\text{interval}}$ & Time interval between two adjacent scans (s) & 0.5 \\
\bottomrule
\label{Radar_parameters}
\end{tabular}
\end{table}
\begin{table}[htbp]
\caption{Object Simulation Settings}
\centering
\begin{tabular}{lcccc}
\toprule
Object & Range (m) & Azimuth ($^{\circ}$) & Radial velocity (m/s) & RCS (dBsm) \\
\midrule
Static IO & 16 / 56.7 & 60 / 40 & 0 & 37 / 60 \\
Mobile IO & 25 & 20 & 10 & 23 \\
Target & 57 & 50 & 0.5 & -20 \\
\bottomrule
\label{Object_parameters}
\end{tabular}
\end{table}
The relevant parameters for Weibull clutter simulation can be found in Table \ref{Weibull_parameters}. As mentioned in \cite{weibullparasetting2012}, the Weibull scale and shape parameter $p$ and $q$ for 24 GHz short range radar are respectively 1.6 and 6.9 for traffic road Weibull clutter. Meanwhile, as mentioned in \cite{jin2015analysis}, for wideband radar, the PSD of the ground clutter is usually described by the all-pole model. Thus, an all-pole model with an exponent $\gamma$, which is equal to 3, is used as the PSD of Weibull clutter in the simulation. The probability density function (PDF) and the PSD of the simulated Weibull clutter can be found in Fig. \ref{Weibull_simulation_verification}, where the amplitude in Fig. \ref{PDF} is normalized by its mean and the PSD in Fig. \ref{PSD} is normalized by its maximum. Accordingly, the correctness of the Weibull clutter simulation can be validated, as the PDF and PSD of the simulated clutter fit well with the theoretical ones. Additionally, we define the clutter power coefficient $\xi$ as the ratio of the clutter power to the weak target echo power, which is equal to 2000 in the simulation. For simplification, complex Additive White Gaussian Noise (AWGN) is considered as the background noise. \par
\begin{table}[htbp]
\caption{Weibull Clutter Simulation Settings}
\centering
\begin{tabular}{lll}
\toprule
Parameters & Description & Value \\
\midrule
$p$ & Scale parameter & 1.6 \\
$q$ & Shape parameter & 6.9 \\
$v_{\text{w}}$ & Radial velocity of wind (m/s) & 5 \\
$\gamma$ & Exponent of the all-pole model & 3 \\
$\xi$ & Clutter power coefficient & 2000\\
\bottomrule
\label{Weibull_parameters}
\end{tabular}
\end{table}
The relevant parameters for CFAR detectors and Godec in the simulation can be found in Table \ref{CFAR_parameters} and Table \ref{Godec_parameters}, respectively. \par
\begin{table}[htbp]
\caption{CFAR Parameters}
\centering
\begin{tabular}{lll}
\toprule
Parameters & Description & Value \\
\midrule
$N_{\text{G}}$ & Number of guard cells & 8 \\
$N_{\text{R}}$ & Number of reference cells & 40 \\
$P_{\text{fa}}$ & False alarm rate & 1e-9 \\
$k$ & The $k$-th strongest reference cell & 10 \\
\bottomrule
\label{CFAR_parameters}
\end{tabular}
\end{table}
\begin{table}[htbp]
\caption{Godec Parameters}
\centering
\begin{tabular}{lll}
\toprule
Parameters & Description & Value \\
\midrule
$r$ & Rank upper bound constraint on the low-rank matrix $\mathbf{L}$ & 1 \\
$\zeta$ & Integer exponent of the power scheme modification & 3 \\
$\epsilon_0$ & The pre-determined error bound & 1e-3 \\
$\delta$ & The threshold for the error difference between two adjacent iterations & 1e-4 \\
$iter_{\text{max}}$ & The maximum iteration times & 100 \\
\bottomrule
\label{Godec_parameters}
\end{tabular}
\end{table}
\subsection{Masking Effects}
\begin{figure}[htbp]
\centerline{\includegraphics[width=0.45\textwidth]{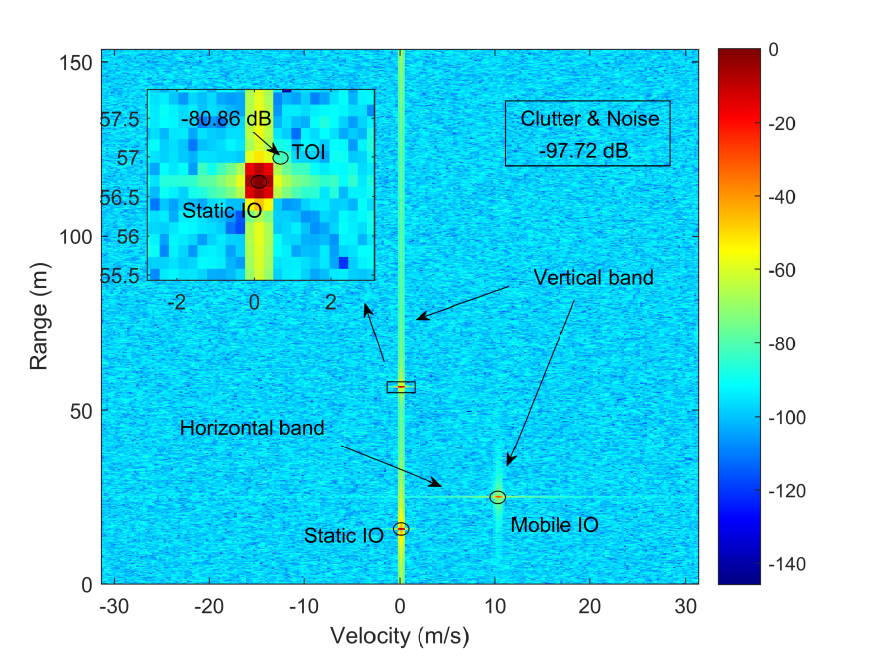}}
\caption{Radar range-velocity map without clutter suppression.}
\label{Raw_RD_map}
\end{figure}

Here we show the results in the first radar scan as an example. As shown in Fig. \ref{Raw_RD_map}, the range-velocity map obtained by the two-dimensional FFT is normalized by its maximum. The intensity of the echoes of TOI is 16.86 dB higher than that of the background clutter and noise. Besides, it can be seen that there are horizontal and vertical bands at the location of the strong IOs in the range-velocity map, which is caused by the spectral leakage during the two-dimensional FFT. As shown in the enlarged part of Fig. \ref{Raw_RD_map}, it is difficult to distinguish the TOI from the spectral extension of the strong IO. \par
\begin{figure}[htbp]
	\centering
    \subfloat[CA-CFAR detection result]{
	\includegraphics[width=0.45\textwidth]{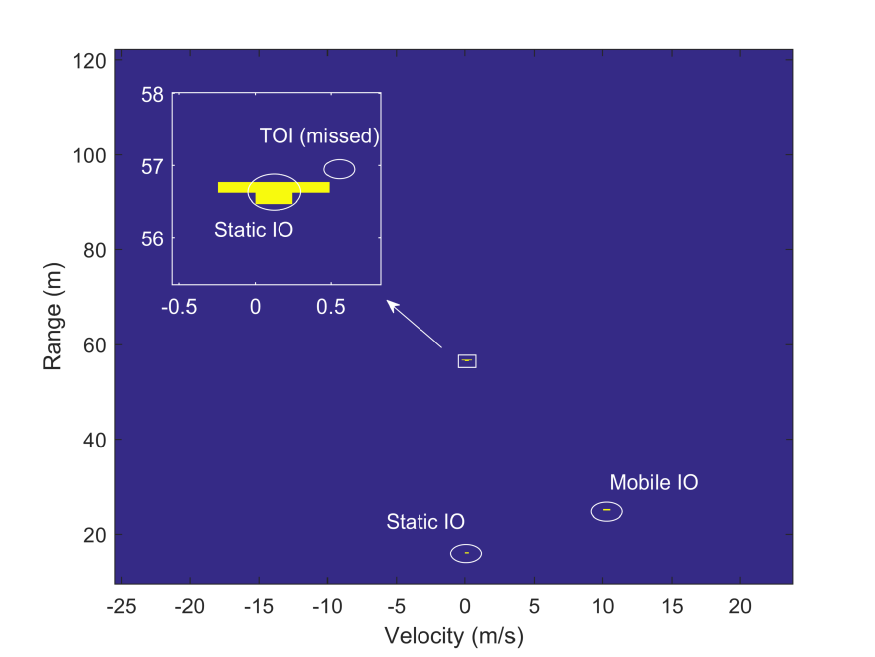}}
    \label{CA-CFAR_only}
    \subfloat[OS-CFAR detection result]{
	\includegraphics[width=0.45\textwidth]{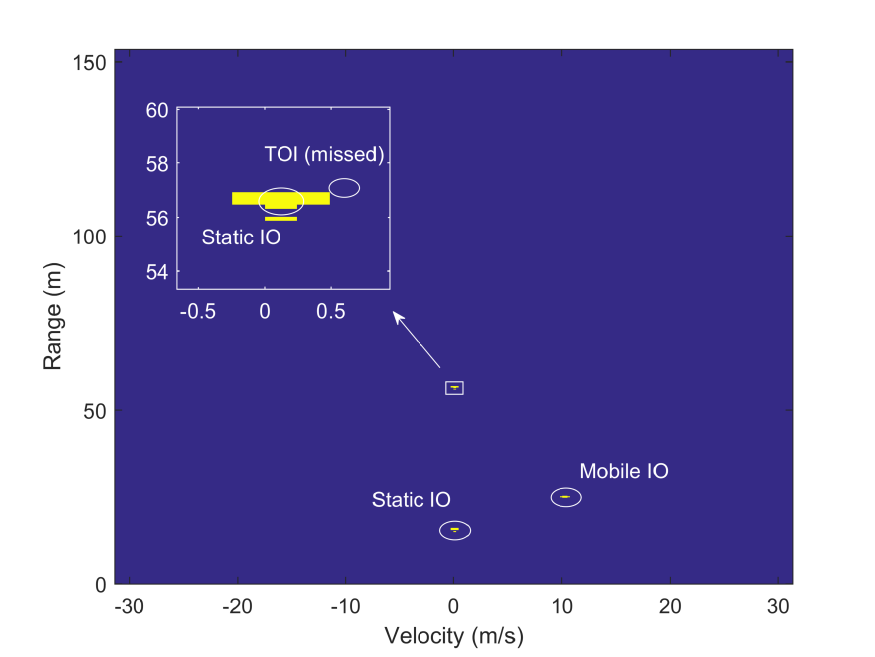}}
    \label{OS-CFAR_only}
	\caption{Masking effects in slow-moving weak target detection.}
	\label{masking_effect}
\end{figure}

In this case, CA-CFAR and OS-CFAR are both performed on Fig. \ref{Raw_RD_map}, and the CFAR detection results can be found in Fig. \ref{masking_effect}. It can be found that in all of the 10 scans, neither CA-CFAR nor OS-CFAR can detect the weak TOI due to the masking effects caused by the nearby strong static IO. \par
\subsection{Clutter Suppression}
In this part, we show the performance of different target detection schemes using MTI and Godec for clutter suppression, and further compare the time consumption of those schemes that are able to detect the slow-moving weak targets.
\subsubsection{MTI}
\begin{figure}[htbp]
\centerline{\includegraphics[width=0.45\textwidth]{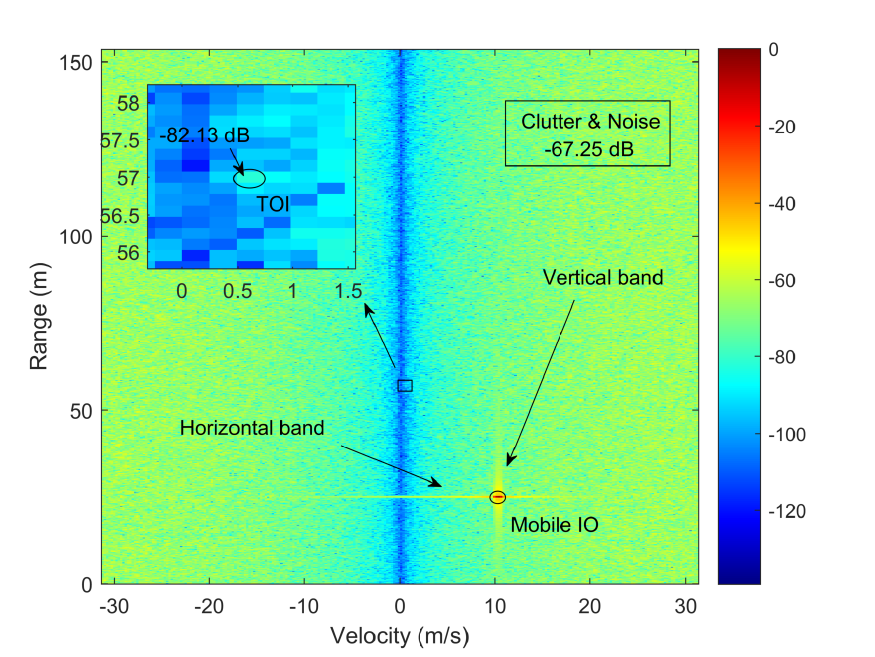}}
\caption{Radar range-velocity map processed by MTI.}
\label{MTI_RD_map}
\end{figure}

As shown in Fig. \ref{MTI_RD_map}, the range-velocity map processed by a single delay-line canceller in the first radar scan is normalized by its maximum. It can be seen from the enlarged part of Fig. \ref{MTI_RD_map} that although the echoes of the strong static IO have been well removed, the echoes of the slow-moving weak target also suffer an intensity loss, whose intensity is even 14.88 dB lower than that of the background clutter and noise. \par
\begin{figure}[htbp]
	\centering
    \subfloat[CA-CFAR detection result]{
	\includegraphics[width=0.45\textwidth]{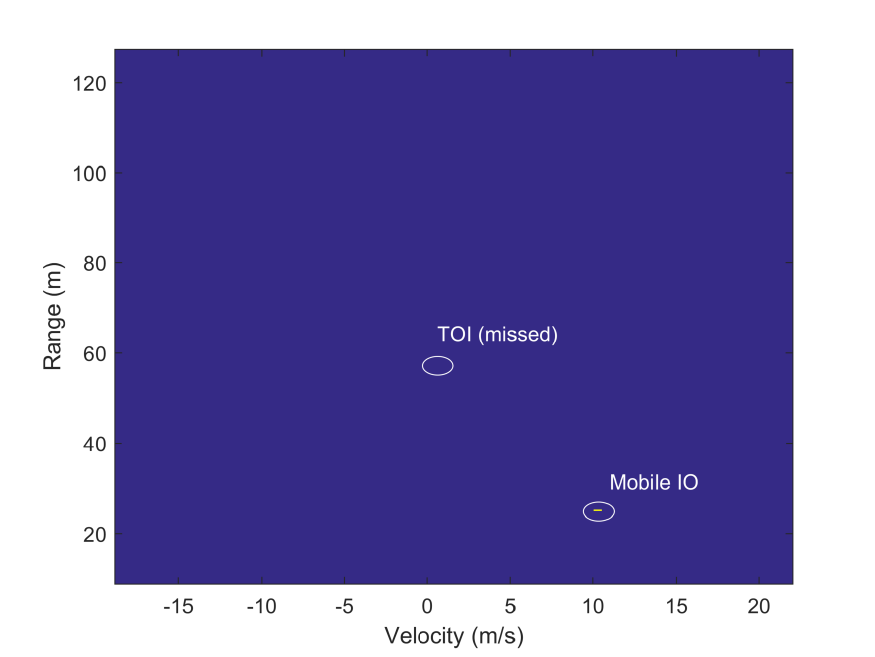}}
    \label{CA-CFAR_only}
    \subfloat[OS-CFAR detection result]{
	\includegraphics[width=0.45\textwidth]{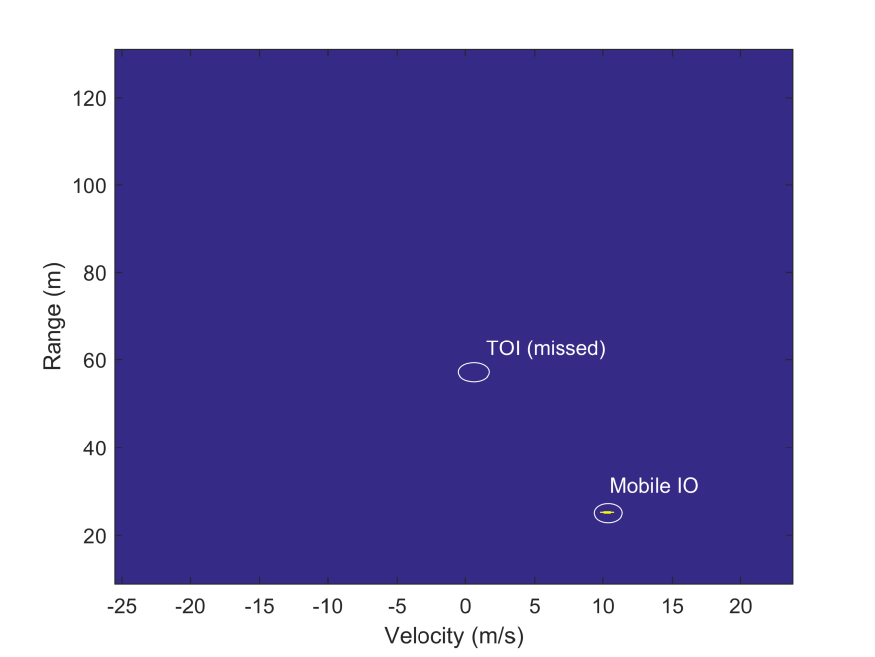}}
    \label{OS-CFAR_only}
	\caption{Detection performance with MTI clutter suppression.}
	\label{MTI_results}
\end{figure}

In this case, CA-CFAR and OS-CFAR are both performed on Fig. \ref{MTI_RD_map}, and the CFAR detection results can be found in Fig. \ref{MTI_results}. Considering all of the 10 scans, the results show that CA-CFAR still cannot detect the weak TOI, but OS-CFAR can obtain the correct detection results in the 8-th scan, i.e., the detection probability of the MTI + OS-CFAR scheme is $10\%$. \par
\subsubsection{Godec}
\begin{figure*}[htbp]
	\centering
	\subfloat[\label{RPCA_static}Static range-velocity map]{
	\includegraphics[width=0.45\textwidth]{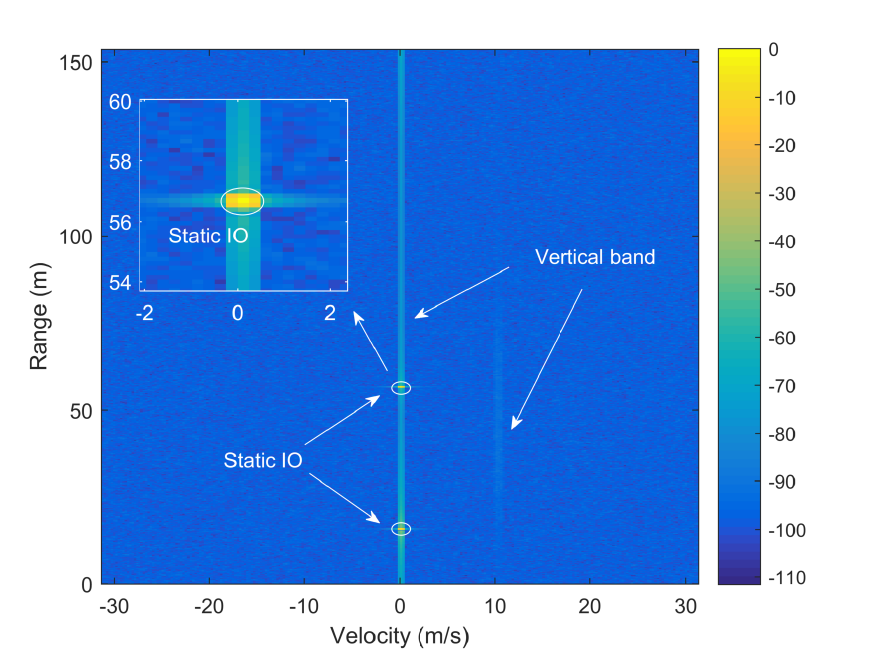}}
    \quad  
    \subfloat[\label{RPCA_mobile}Dynamic range-velocity map]{
	\includegraphics[width=0.45\textwidth]{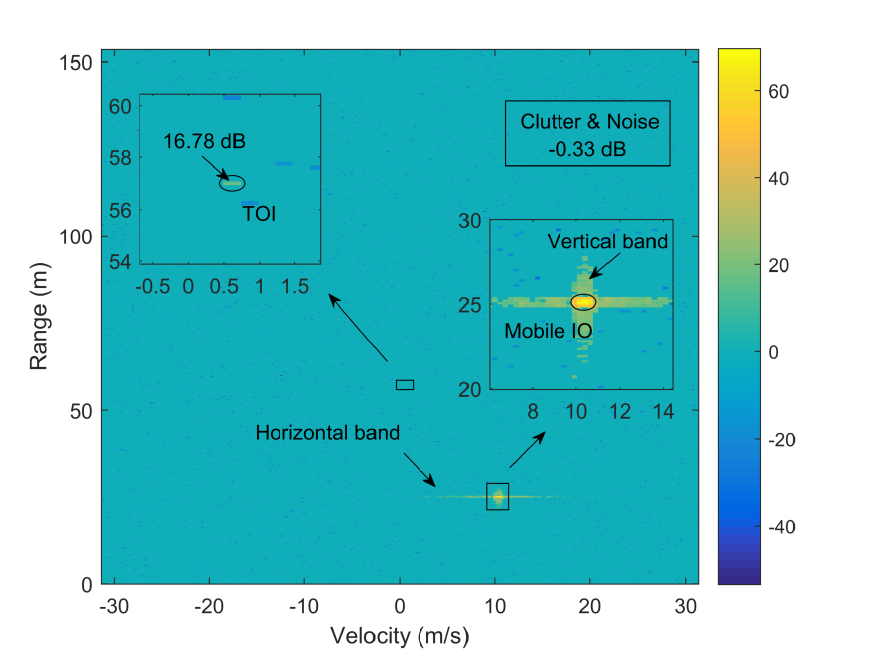}}
    \quad
    \subfloat[\label{RPCA_CA-CFAR}CA-CFAR detection result]{
	\includegraphics[width=0.45\textwidth]{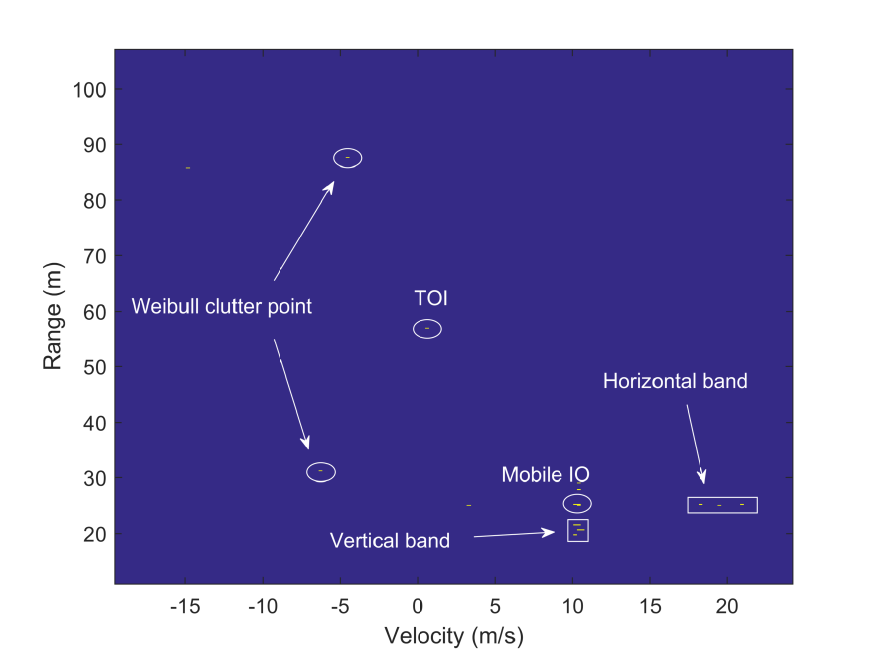}}
    \quad  
	\caption{Results of the Godec + CA-CFAR scheme with $N_{\text{mov}} = 18$.}
	\label{RPCA_static_mobile}
\end{figure*}

As shown in Fig. \ref{RPCA_static_mobile}, Godec achieves suppression of the strong static IOs by separating cells with constant and varying intensity in multiple scans into different range-velocity maps. Besides the Godec parameters shown in Table \ref{Godec_parameters}, there is still a hyperparameter $N_{\text{mov}}$ which determines the sparsity upper bound constraint on the sparse matrix $\mathbf{S}$. Here we assume that $N_{\text{mov}} = 18$, and it can be seen from the enlarged part of Fig. \ref{RPCA_static} that the echoes of TOI nearby the strong static IO are well separated. Besides, in Fig. \ref{RPCA_mobile}, the cell on the position of TOI is highlighted, whose intensity can achieve 17.11 dB higher than that of the background clutter and noise. Compared with the intensity loss caused by MTI, it can be seen that the proposed Godec-based method can effectively preserve the echoes of TOI from being suppressed. \par
Considering the computational complexity, OS-CFAR is not considered here. CA-CFAR is performed on Fig. \ref{RPCA_mobile}, whose result is shown in Fig. \ref{RPCA_CA-CFAR}, where the TOI can be successfully detected. Considering all of the 10 scans, it can be found that the detection probability of the Godec + CA-CFAR scheme can reach up to $100\%$. \par
\subsubsection{Time Consumption Comparison}
For the MTI + OS-CFAR and the Godec + CA-CFAR schemes which can both detect the TOI, here we give the average running time of ten running tests in Table \ref{time_comparison}. It can be calculated that the time consumption of the Godec + CA-CFAR scheme is $93.26\%$ of that of the MTI + OS-CFAR scheme. However, the time consumption accounted for in the Godec + CA-CFAR scheme does not include the time required to determine the optimal hyperparameter $N_{\text{mov}}$. Consequently, in practical applications, the overall time consumption of the Godec + CA-CFAR scheme is expected to exceed that of the MTI + OS-CFAR scheme. This indicates that the Godec + CA-CFAR scheme prioritizes achieving high reliability in detecting the slow-moving weak targets at the cost of increasing the time complexity. \par
\begin{table}[htbp]
\caption{Average Time Consumption}
\centering
\begin{tabular}{lcccc}
\toprule
Process & MTI & Godec & CA-CFAR & OS-CFAR \\
\midrule
Time (s) & 0.0414 & 0.3378 & 7.6389 & 8.5115 \\
\bottomrule
\label{time_comparison}
\end{tabular}
\end{table}
\subsubsection{Tradeoffs in the Godec-based Solution}
When searching for the optimal hyperparameter $N_{\text{mov}}$, three repeated runs are conducted for each value of $N_{\text{mov}}$, as the Weibull clutter generated in the simulation varies with each run. The average number of the false alarm cells in 10 scans $N_{\text{fa}}$, the detection probability of the TOI $P_{\text{d}}$, the iteration times required for convergence $iter_{\text{cov}}$ and the average performance function in three runs $\overline{f}$ obtained by the Godec + CA-CFAR scheme with different $N_{\text{mov}}$ are presented in Table \ref{tradeoff_curve}. From this, the following observations can be made:
\begin{itemize}
\item[$\bullet$] Generally, as $N_{\text{mov}}$ increases, $N_{\text{fa}}$ also increases, with a slight decrease observed after $N_{\text{mov}} = 100$. The detection probability $P_{\text{d}}$ reaches its maximum in the range of $N_{\text{mov}} = 25$ to $30$. Beyond this range, further increase in $N_{\text{mov}}$ lead to a slight decrease in $P_{\text{d}}$, and it generally remains above $70\%$. Additionally, the number of iteration times required for convergence $iter_{\text{cov}}$ increases as $N_{\text{mov}}$ grows.
\item[$\bullet$] In the table, two tradeoffs can be observed. When $N_{\text{mov}} \leq 30$, $iter_{\text{cov}}$ is always equal to $2$, and the tradeoff primarily lies between $N_{\text{fa}}$ and $P_{\text{d}}$. However, when $N_{\text{mov}} > 100$, $P_{\text{d}}$ changes slightly, and the tradeoff shifts to being mainly between $N_{\text{fa}}$ and $iter_{\text{cov}}$.
\item[$\bullet$] In the table, as $N_\text{mov}$ increases, $\overline{f}$ exhibits a trend of first increasing and then decreasing in general, and it reaches its maximum when $N_\text{mov} = 30$.
\end{itemize}
\begin{table}[htbp]
\caption{$N_{\text{fa}}$, $P_{\text{d}}$, $iter_{\text{cov}}$ and $\overline{f}$ with Different $N_{\text{mov}}$}
\centering
\begin{tabular}{lccccccccccccc}
\toprule
$N_{\text{mov}}$ & 1 & 5 & 8 & 11 & 14 & 18 & 21 & 25 & 30 & 40 & 50 & 100 & 200 \\
\midrule
\multirow{3}{*}{$N_{\text{fa}}$} & 3.4 & 5.5 & 4.8 & 5.1 & 6.8 & 13.1 & 13.8 & 15.3 & 17.4 & 20.3 & 21.9 & 34.5 & 27.1 \\
 & 3.4 &5.8 &5.7 &6.0 &7.5	&10.2	&14.1	&13.9	&14.4	&20.1	&23.8	&34.4	&26.4 \\
 & 2.9 &4.6 &5.7 &6.4 &8.3	&10.3	&12.1	&13.2	&14.6	&19.2	&24.3	&34.8	&25.2 \\
\midrule
\multirow{3}{*}{$P_{\text{d}}$} & 0	&0.2	&0.2	&0.4 &0.6	&0.6	&0.9	&1.0	&0.8	&0.9	&0.8	&0.9	&0.8 \\
& 0	&0.3	&0.4	&0.4 &0.5	&0.7	&0.6	&0.7	&0.9	&0.8	&0.7	&0.5	&0.7 \\
& 0	&0.1	&0.1	&0.4 &0.5	&1.0	&0.7	&0.7	&0.8	&0.7	&0.9	&0.8	&0.7 \\
\midrule
\multirow{3}{*}{$iter_{\text{cov}}$} &2	&2	&2	&2 &2	&2	&2	&2	&2	&3	&3	&4	&5 \\
& 2	&2	&2	&2 &2	&2	&2	&2	&2	&3	&3	&4	&5 \\
& 2	&2	&2	&2 &2	&2	&2	&2	&2	&3	&3	&4	&5 \\
\midrule
$\overline{f}$ &0.309 	&0.340 	&0.348 	&0.387 	&0.413 	&0.462 	&0.451 	&0.467 	&0.474 	&0.379 & 0.377		&0.315 	&0.293 \\
\bottomrule
\end{tabular}
\label{tradeoff_curve}
\end{table}
For the observations above, the following interpretations are provided:
\begin{itemize}
\item[$\bullet$] For $N_{\text{fa}}$ and $P_{\text{d}}$: The increase in $N_{\text{mov}}$ causes the intensity threshold for extracting sparse components from the range-velocity map to decrease. As a result, more dynamic cells are extracted, leading to an increase in $N_{\text{fa}}$ and $P_{\text{d}}$. However, when $N_{\text{mov}}$ becomes sufficiently large ($N_{\text{mov}} > 100$), Godec is able to extract the sparse components caused by the Weibull clutter, which leads to a huge number of clutter cells appearing in the dynamic range-velocity map as shown in Fig. \ref{RPCA_mobile}. During CFAR detection, these clutter cells may have the masking effects on each other, which results in the decrease in $N_{\text{fa}}$. Moreover, the numerous clutter cells also cause a reduction in $P_{\text{d}}$ to some extent.
\item[$\bullet$] For $iter_{\text{cov}}$: The noise component $\mathbf{N}$ with large magnitude will decelerate the convergence process \cite{zhou2011godec}. Therefore, when $N_{\text{mov}}$ is small ($N_{\text{mov}} \leq 30$), the extracted sparse components do not yet include the dynamic cells caused by the Weibull clutter. In this case, the intensity of $\mathbf{N}$ is relatively small compared to that of the extracted sparse components, resulting in fewer iteration times for convergence. On the other hand, when $N_{\text{mov}}$ is large ($N_{\text{mov}} > 30$), the extracted sparse components not only include the TOI but also a portion of dynamic cells caused by the Weibull clutter. In this case, the intensity of $\mathbf{N}$ becomes more significant relative to these dynamic cells, leading to an increase in $iter_{\text{cov}}$ as $N_{\text{mov}}$ increases.
\item[$\bullet$] For selecting the optimal $N_{\text{mov}}$, to make the results more robust, here we consider that any $N_\text{mov}$ that result in $f$ reaching $95\%$ of its maximum are acceptable. Accordingly, the recommended range for $N_\text{mov}$ is approximately between 18 and 30 in the simulation.
\end{itemize}

\section{Experiment Validation}
\subsection{Overview}
\begin{figure}[htbp]
\centerline{\includegraphics[width=0.45\textwidth]{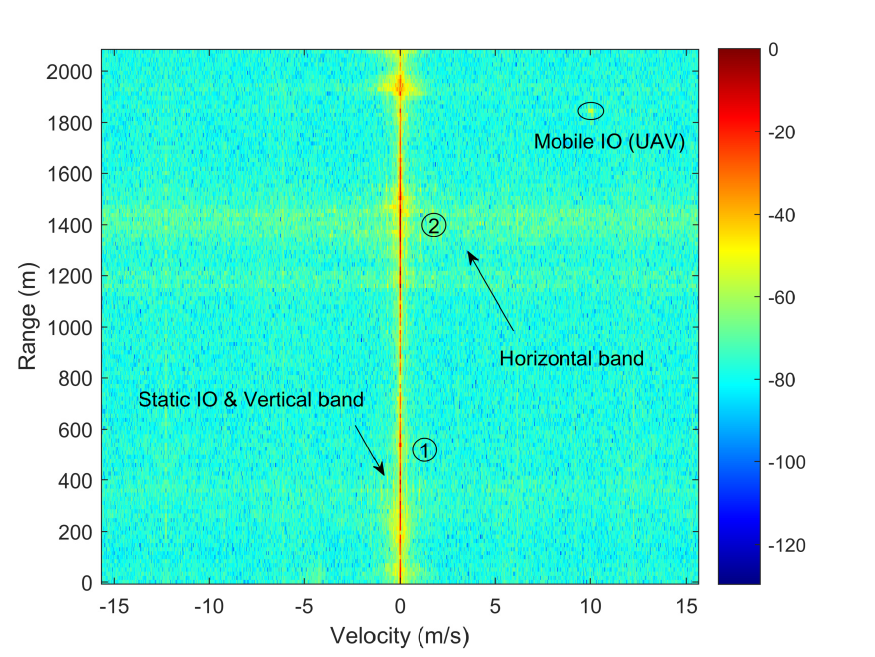}}
\caption{Range-velocity map obtained by realistic measurement.}
\label{Experiment_RD_map}
\end{figure}

The data analyzed in this section are provided by a private company specializing in radar target detection, under a confidentiality agreement. Due to the proprietary nature of the data, the exact identity of the company and further details cannot be disclosed. Here, we present one of the realistic range-velocity maps, as shown in Fig. \ref{Experiment_RD_map}. After experiment validation, it was found that there is only one weak mobile IO (UAV) moving at a constant speed of 10 m/s, while the other components are either the static IOs or the uninterested clutter. \par
Due to limitations in the experimental conditions, we introduce a TOI aligning with the assumptions made in this article into the range-velocity map by simulation. The echo intensity of the simulated TOI is approximately equal to that of the realistic weak UAV on the same range. Generally, as shown in Fig. \ref{Experiment_RD_map}, the slow-moving weak target detection can be considered from these two cases: one in which the TOI is beyond the horizontal band, and the other in which it is within the horizontal band. In the following parts of this section, we will evaluate the performance of the proposed Godec-based solution in detecting the TOI in these two cases.
\subsection{Case 1: TOI Located beyond the Horizontal Band}
\begin{figure}[htbp]
\centerline{\includegraphics[width=0.45\textwidth]{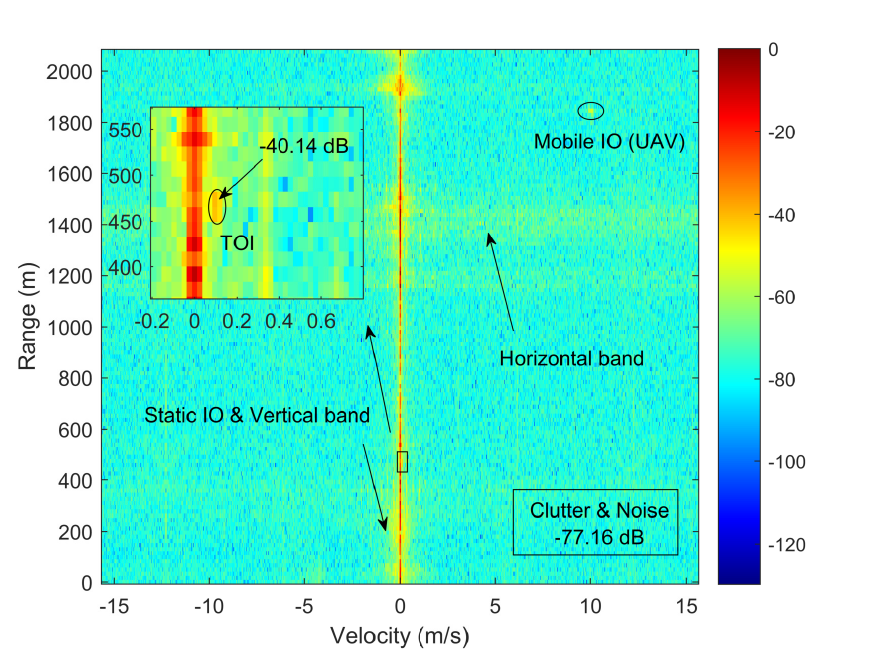}}
\caption{Range-velocity map with TOI located beyond the horizontal band.}
\label{Experiment_RD_map_1}
\end{figure}

\begin{figure}[htbp]
	\centering
    \subfloat[CA-CFAR detection result]{
	\includegraphics[width=0.45\textwidth]{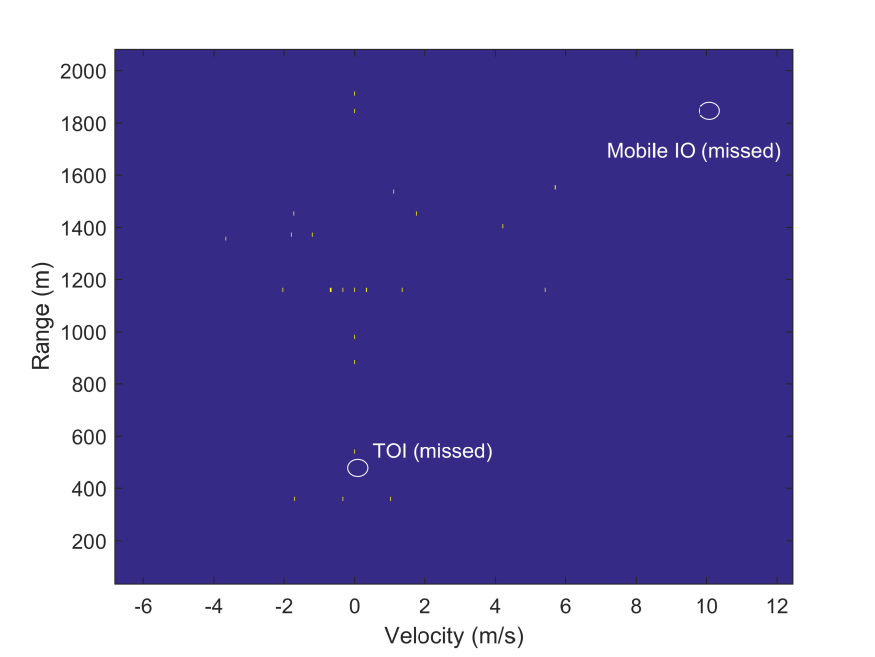}}
    \label{CA-CFAR_only_exp}
    \subfloat[OS-CFAR detection result]{
	\includegraphics[width=0.45\textwidth]{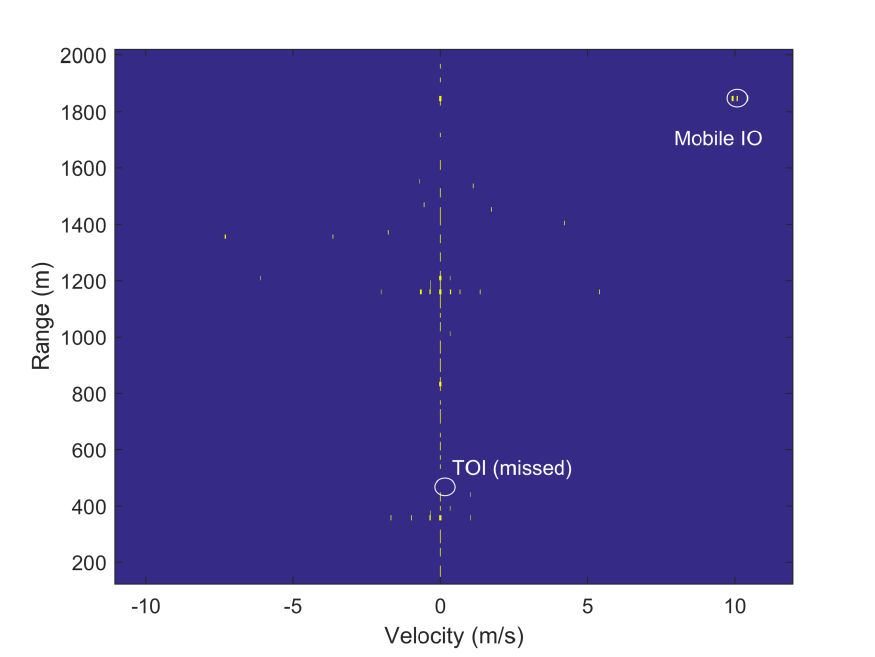}}
    \label{OS-CFAR_only_exp}
	\caption{Masking effects in experiment.}
	\label{masking_effect_exp}
\end{figure}

Taking scan 1 as an example, as shown in Fig. \ref{Experiment_RD_map_1}, the simulated TOI is located at a distance of 468 m from the radar, with a normalized intensity of approximately -40 dB and a velocity at 0.1 m/s. The results of applying CFAR detection directly to the range-velocity map without clutter suppression is shown in Fig. \ref{masking_effect_exp}. Considering all 10 scans, the detection probabilities of CA-CFAR and OS-CFAR are $0$ and $10\%$, respectively. \par
\begin{figure}[htbp]
\centerline{\includegraphics[width=0.45\textwidth]{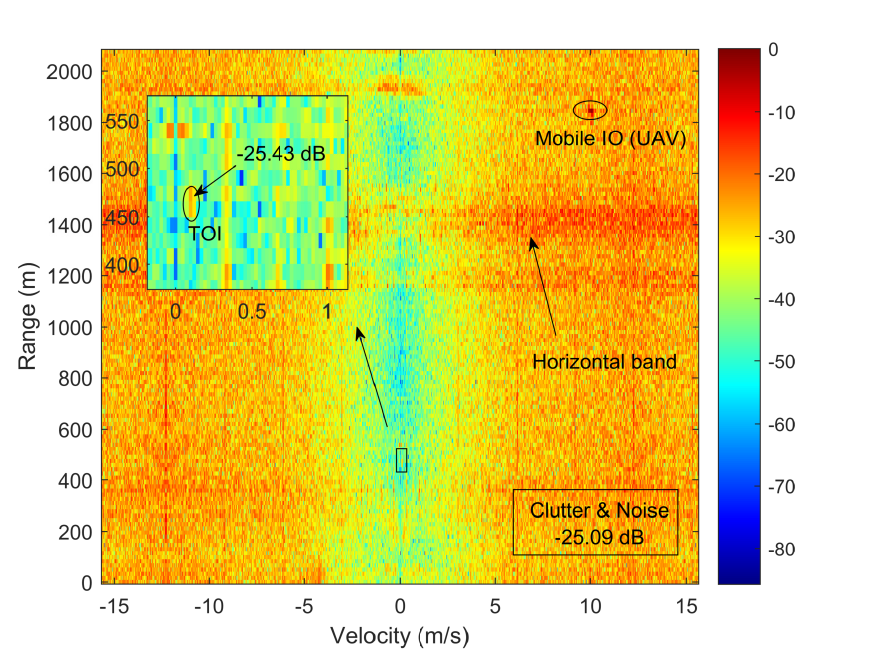}}
\caption{Realistic range-velocity map processed by MTI.}
\label{MTI_RD_map_exp}
\end{figure}

\begin{figure}[htbp]
	\centering
    \subfloat[CA-CFAR detection result]{
	\includegraphics[width=0.45\textwidth]{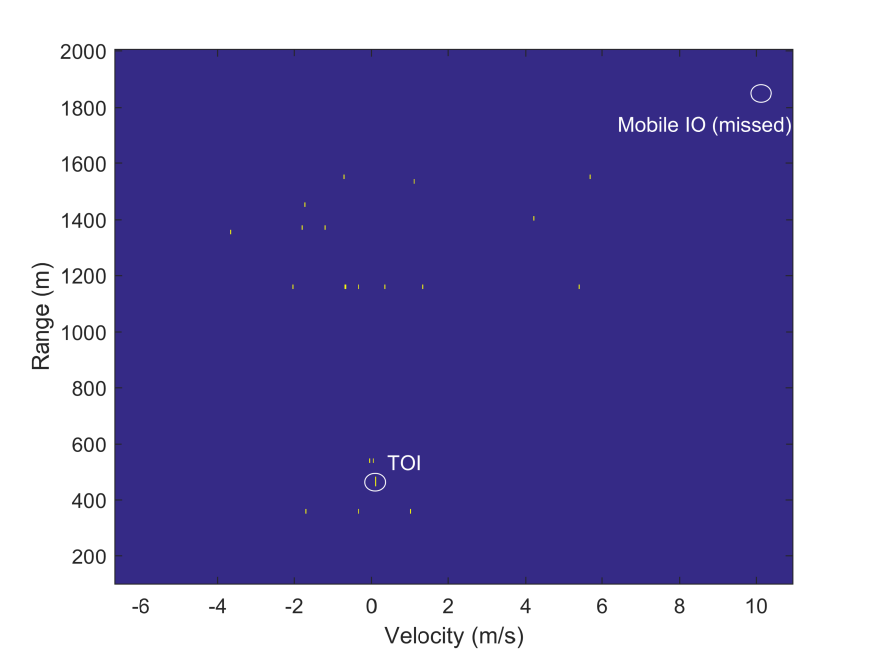}}
    \label{CA-CFAR_only}
    \subfloat[OS-CFAR detection result]{
	\includegraphics[width=0.45\textwidth]{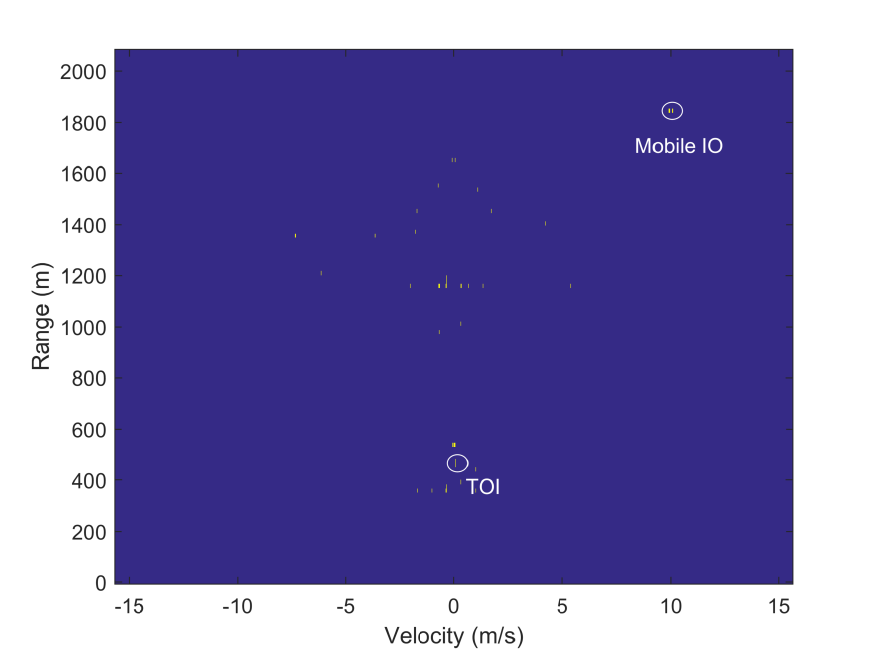}}
    \label{OS-CFAR_only}
	\caption{Realistic detection performance with MTI clutter suppression.}
	\label{MTI_results_exp}
\end{figure}

The realistic range-velocity map processed by MTI is shown in Fig. \ref{MTI_RD_map_exp}, and the CFAR detection results can be seen in Fig. \ref{MTI_results_exp}. In this case, the detection probabilities of CA-CFAR and OS-CFAR are respectively $50\%$ and $90\%$, and the performances of the detection schemes mentioned above are summarized in Table \ref{schemes_summ}. \par
\begin{figure*}[htbp]
	\centering
	\subfloat[\label{RPCA_static_exp}Static range-velocity map]{
	\includegraphics[width=0.45\textwidth]{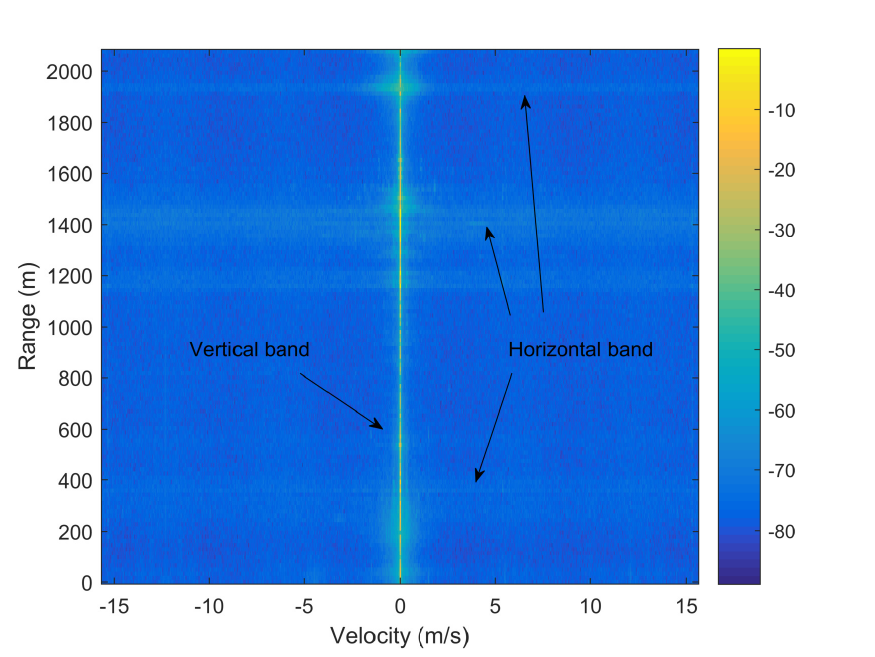}}
    \quad  
    \subfloat[\label{RPCA_mobile_exp}Dynamic range-velocity map]{
	\includegraphics[width=0.45\textwidth]{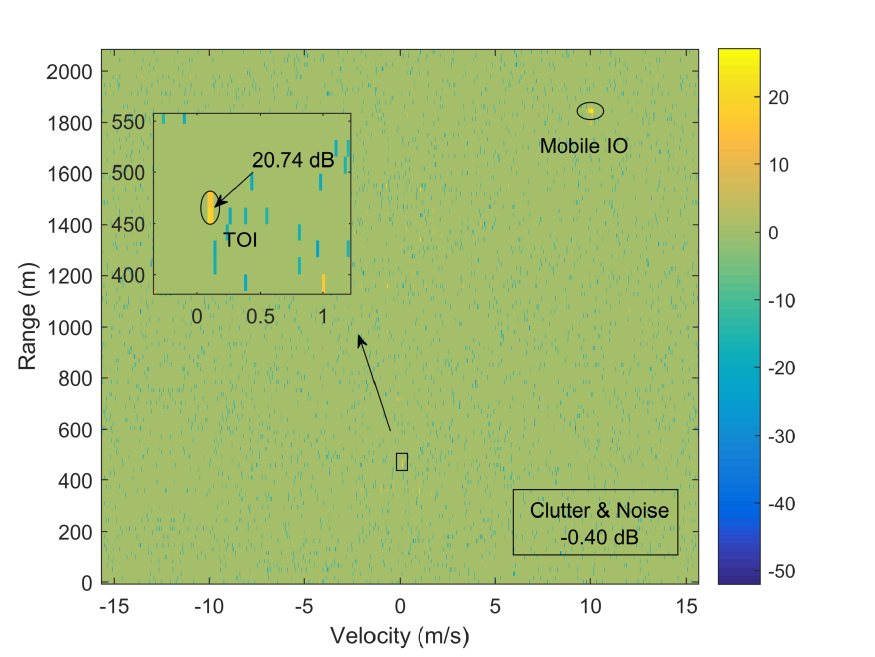}}
    \quad
    \subfloat[\label{RPCA_CA-CFAR_exp}CA-CFAR detection result]{
	\includegraphics[width=0.45\textwidth]{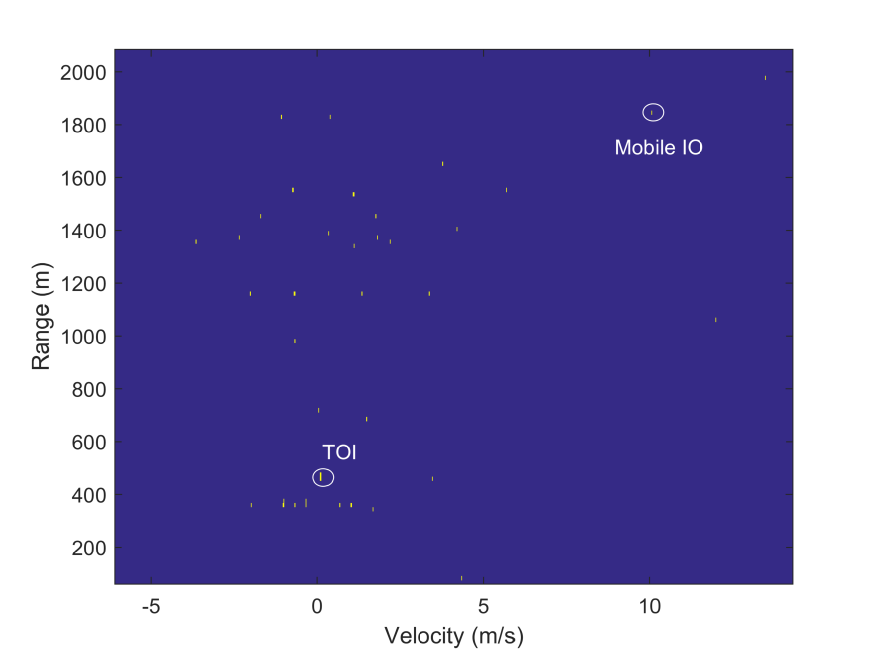}}
    \quad  
	\caption{Experiment results of the Godec + CA-CFAR scheme with $N_{\text{mov}} = 3$.}
	\label{RPCA_static_mobile_exp}
\end{figure*}

\begin{table}[htbp]
\caption{Detection Performances of the Typical Schemes}
\centering
\begin{tabular}{lcccc}
\toprule
Scheme & CA-CFAR only & OS-CFAR only & MTI + CA-CFAR & MTI + OS-CFAR \\
\midrule
$P_{\text{d}}$ & $0$ & $10\%$ & $50\%$ & $90\%$ \\
$N_{\text{fa}}$& $24.7$ & $115.1$ & $21.7$ & $42.6$ \\
\bottomrule
\end{tabular}
\label{schemes_summ}
\end{table}
For the proposed Godec-based solution, here we assume $N_{\text{mov}} = 3$, and the static and dynamic range-velocity maps obtained are presented as Fig. \ref{RPCA_static_exp} and Fig. \ref{RPCA_mobile_exp}. After CA-CFAR detection shown in Fig. \ref{RPCA_CA-CFAR_exp}, the detection probability of TOI can reach $100\%$, with the average number of the false alarm cells in 10 scans $N_{\text{fa}} = 92.7$. Besides, the time consumption comparison of the Godec-based scheme and the MTI + OS-CFAR scheme can be found in Table \ref{time_comparison_exp}, and the tradeoffs revealed in the simulation section can be validated from Table \ref{tradeoff_curve_exp_1}, where we can recommend the range for $N_{\text{mov}}$ from 2 to 4. It can be observed that although the Godec-based scheme achieves a $10\%$ improvement in the detection probability, this comes at the cost of a doubling of false alarm cells and a significantly higher time consumption. Therefore, when the TOI is beyond the horizontal band, we may recommend using the MTI-based scheme. \par
\begin{table}[htbp]
\caption{Average Time Consumption in Experiment}
\centering
\begin{tabular}{lcccc}
\toprule
Process & MTI & Godec & CA-CFAR & OS-CFAR \\
\midrule
Time (s) & 0.0199 & 0.2402 & 4.7611 & 5.3720 \\
\bottomrule
\label{time_comparison_exp}
\end{tabular}
\end{table}
\begin{table}[htbp]
\caption{$N_{\text{fa}}$, $P_{\text{d}}$, $iter_{\text{cov}}$ and $f$ with Different $N_{\text{mov}}$ in Case 1}
\centering
\begin{tabular}{lccccccccccc}
\toprule
$N_{\text{mov}}$ & 1 & 2 & 3 & 4 & 5 & 7 & 10 & 15 & 25 & 40 & 70 \\
\midrule
$N_{\text{fa}}$ &14.5	&49.3	&92.7	&93.5	&86.0	&73.2	&70.0	&58.5	&47.7	&32.0	&13.6 \\
$P_{\text{d}}$ &0.3	&0.9	&1.0	&1.0	&0.9	&0.9	&0.9	&0.9	&0.9	&0.9	&0.9 \\
$iter_{\text{cov}}$ &2	&2	&2	&2	&3	&3	&4	&4	&5	&7	&9 \\
$f$ &0.341 	&0.480 	&0.503 	&0.503 	&0.395 	&0.395 	&0.354 	&0.354 	&0.330 	&0.304 	&0.298 \\
\bottomrule
\end{tabular}
\label{tradeoff_curve_exp_1}
\end{table}
\subsection{Case 2: TOI Located within the Horizontal Band}
\begin{figure}[htbp]
\centerline{\includegraphics[width=0.45\textwidth]{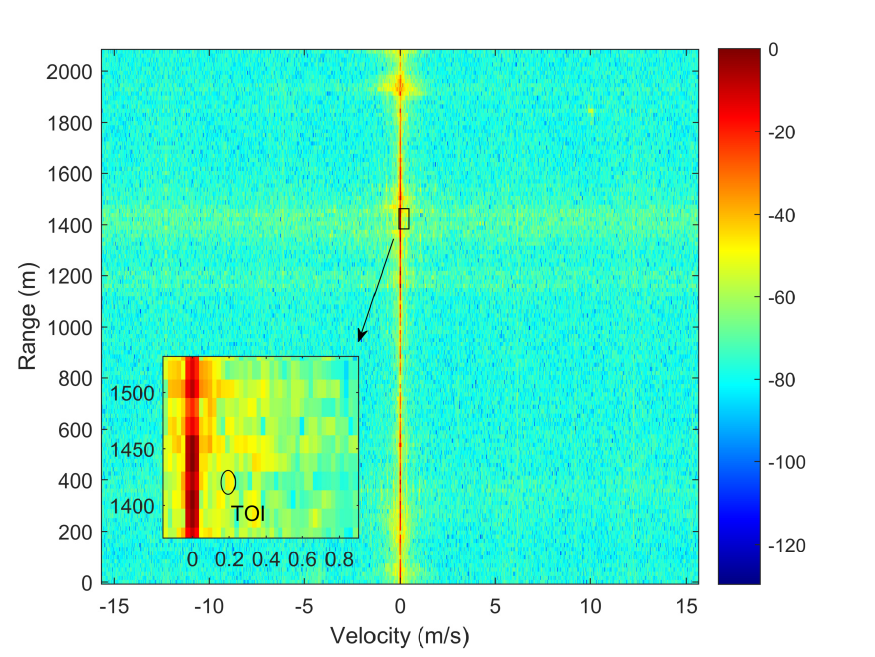}}
\caption{Range-velocity map with TOI located within the horizontal band.}
\label{Experiment_RD_map_2}
\end{figure}

In this case, we adjusted the position of the simulated TOI to a distance of 1418 m from the radar, with a normalized intensity of approximately -45 dB and a velocity of 0.1 m/s. By searching for the optimal $N_{\text{mov}}$, we found that regardless of the value of $N_{\text{mov}}$, the detection probability of the TOI was 0. Therefore, we reset the velocity of TOI to 0.2 m/s, and the range-velocity map is shown in Fig. \ref{Experiment_RD_map_2}, where we can find that the TOI is located within the horizontal band. \par
In this case, the MTI + OS-CFAR scheme fails to detect the TOI in any scan, with the average number of false alarm cells across 10 scans being 42.1. Besides, the performance of the Godec-based scheme is shown in Table \ref{tradeoff_curve_exp_2}. It can be observed that the detection probability can reach up to $40\%$ when $N_{\text{mov}} = 3$ and $4$, proving the reliability of the proposed solution on slow-moving weak target detection. Therefore, compared to case 1 mentioned above, we may recommend using the proposed Godec-based scheme when the TOI is within the horizontal band. \par
\begin{table}[htbp]
\caption{$N_{\text{fa}}$, $P_{\text{d}}$, $iter_{\text{cov}}$ and $f$ with Different $N_{\text{mov}}$ in Case 2} 
\centering
\begin{tabular}{lcccccccccc}
\toprule
$N_{\text{mov}}$ & 1 & 2 & 3 & 4 & 5 & 6 & 7 & 8 & 10 & 15 \\
\midrule
$N_{\text{fa}}$ &14.4	&48.2	&91.4	&91.9	&83.5	&79.8	&71.5	&66.3	&68.7	&57.0 \\
$P_{\text{d}}$ &0	&0.2	&0.4	&0.4	&0.3	&0.3	&0.3	&0.3	&0.2	&0.1 \\
$iter_{\text{cov}}$ &2	&2	&2	&2	&3	&3	&3	&3	&4	&4 \\
$f$ &0.266 	&0.305 	&0.353 	&0.353 	&0.245 	&0.245 	&0.245 	&0.245 	&0.179 	&0.154 \\
\bottomrule
\end{tabular}
\label{tradeoff_curve_exp_2}
\end{table}
\section{Conclusion}
In this paper, we mainly focus on the reliable clutter suppression for slow-moving weak target radar detection. Specific system settings and the signal processing of LFMCW radar are provided. Besides, a Godec-based method is proposed to suppress the strong echoes of the static IOs. The original Godec algorithm is extended by introducing additional parameters to make it suitable for its application in slow-moving weak target radar detection. To address the challenge of selecting optimal Godec parameters, we employ a piecewise approach to reduce the traversal times required for the searching process, and the optimal parameter values are determined based on a normalized performance function. Simulation results show that, in the presence of masking effects, the proposed Godec-based solution provides a reliable detection performance at a cost of the computational complexity. Furthermore, there are two tradeoffs between the number of false alarm cells, detection probability, and the iteration times required for convergence, which should be considered when selecting the optimal sparsity parameter for Godec. Additionally, experiment validation of the proposed solution verifies the simulation results and further identifies the applicable scenarios for the Godec-based solution, enhancing its practical applicability. \par

\bibliographystyle{IEEEtran}
\bibliography{clutter_suppression}

\end{document}